\documentclass[a4paper,usenatbib]{mnras}
\usepackage{graphicx}
\usepackage[fleqn]{amsmath}
\usepackage[T1]{fontenc}
\usepackage{aecompl}
\usepackage{times}
\usepackage{ae,aecompl}
\usepackage{amssymb,amsfonts,hyperref,color}
\usepackage{ wasysym }

\newcommand\be{\begin{equation}}
\newcommand\en{\end{equation}}

\newcommand\etal{{\rm et al}.\ }

\title[Young Planets]{Directly Detecting the Envelopes of Low-mass Planets Embedded In Protoplanetary Discs and The Case For TW Hydrae}
\author[Z.~Zhu \etal]{
Zhaohuan Zhu$^{1,2}$\thanks{E-mail: zhaohuan.zhu@unlv.edu}, Avery Bailey$^{1,2}$, Enrique Mac\'ias$^{3}$, Takayuki Muto$^{4}$,\newauthor and Sean M. Andrews$^{5}$ \\
$^{1}$Department of Physics and Astronomy, University of Nevada, Las Vegas, 4505 S.~Maryland Parkway, Las Vegas, NV~89154, USA\\
$^{2}$ Nevada Center for Astrophysics, University of Nevada, Las Vegas, 4505 South Maryland Parkway, Las Vegas, NV 89154, USA \\
$^{3}$ ESO Garching, Karl-Schwarzschild-Str. 2, 85748, Garching bei Munchen, Germany \\
$^{4}$ Division of Liberal Arts, Kogakuin University, 1-24-2 Nishi-Shinjyuku, Shinjyuku-ku, Tokyo 163-8677, Japan \\
$^{5}$ Center for Astrophysics | Harvard \& Smithsonian, 60 Garden St., Cambridge, MA 02138, USA \\
}

\date{In original form \today}

\begin{document}
\label{firstpage}
\pagerange{\pageref{firstpage}--\pageref{lastpage}} 
\maketitle

\begin{abstract}
Despite many methods developed to find  young massive planets in protoplanetary discs, it is challenging to  directly detect low-mass planets that are embedded in discs. On the other hand, the core-accretion
theory suggests that there could be a large population of embedded low-mass
young planets at the Kelvin-Helmholtz (KH) contraction phase.  We adopt both 1-D models and 3-D simulations 
to calculate the envelopes around low-mass cores (several to tens of $M_{\oplus}$) with different luminosities, and derive their
thermal fluxes at radio wavelengths. We find that, when the background disc is optically thin at radio wavelengths, radio observations can see through the disc and
probe the denser envelope within the planet's Hill sphere. 
When the
optically thin disc is observed with the resolution reaching one disc scale
height, the radio thermal flux from the planetary envelope around a 10
M$_{\oplus}$ core is more than 10\% higher than the flux from the background
disc. 
The emitting region can be extended and elongated. 
Finally, our model suggests that the au-scale clump at 52 au in the TW Hydrae disc revealed by ALMA 
is consistent with the envelope of an embedded 10-20 $M_{\oplus}$ planet, which can explain the detected flux, the spectral index dip, and the tentative spirals. The observation is 
also consistent with the planet undergoing pebble accretion.
Future ALMA
and ngVLA observations may directly reveal more such low-mass planets, enabling us to study core growth and even reconstruct the planet formation history using the embedded ``protoplanet'' population.

\end{abstract}


\section{Introduction}
Both direct and indirect methods have been proposed to find young planets in protoplanetary discs. We can use high contrast optical/infrared observations
to directly detect them (e.g. PDS 70bc, \citealt{Kepler2018,Haffert2019}), or use the planet's perturbation to the protoplanetary disc to indirectly reveal them.
It could be the velocity perturbation probed by ALMA molecular line observations, such as the ``velocity kink'' from the planet \citep{Pinte2018,Izquierdo2021} 
or sub/super-Keplerian motion at the edges of the planet-induced gaseous gaps \citep{Teague2018, Teague2019}. However, to produce the velocity
perturbation that is detectable by ALMA, the planet needs to be more than 1-2 Jupiter masses \citep{Rabago2021}. To detect lower mass planets, we need to use other disc features,
such as spiral arms (e.g. \citealt{Bae2021, Speedie2022}) or dusty gaps (e.g. \citealt{Zhang2018}). So far, the lowest planet mass we can probe is $\sim$ Neptune mass using the
dusty gaps from ALMA continuum observations \citep{Zhang2018}. 
These Neptune mass planets open very shallow gaps, which can still trap dust particles to be observable. On the other hand,
these indirect methods using disc features cannot pinpoint the planet's location, and the disc features (e.g. spirals and gaps) could also be induced by mechanisms that are
not related to planets at all. 

Exoplanet demographics suggests that the most abundant planets are super-Earths and mini-Neptunes \citep{WinnFabrycky2015}, 
less massive than the detectable planets in protoplanetary discs mentioned above. Furthermore, from a theoretical perspective, many planets will remain in the super-Earth/mini-Neptune mass regime even after the protoplanetary discs dissipate, since
the Kelvin-Helmholtz (KH) contraction phase preceding the run-away accretion phase is the bottleneck of giant planet formation and this phase could be longer than the disc's life time \citep{Pollack1996}. This phase is long since the planet needs to cool to grow. In the traditional core-accretion model, the critical core mass, above which the planet can undergo run-away accretion within the disc's lifetime, is $\sim$10 $M_{\oplus}$ \citep{Mizuno1980,Stevenson1982}. This critical core mass can change depending on the local disc condition (e.g. distance to the star, surface density, disc temperature, \citealt{Piso2014,Lee2014}), and the detailed physical processes around the core  (e.g. \citealt{Brouwers2020}). 

Thus, both observations and theory suggest that there should be a large population of low-mass planets (less than the Neptune mass) embedded in protoplanetary discs. Probing such population could be the key to understanding planet formation. If we can constrain the protoplanet population during this phase and compare it with the population of more massive planets (e.g. giant planets in discs), we can potentially reconstruct the planet evolutionary history. This is similar to constraining protostellar history using the relative number of Young Stellar Objects (YSOs) in the embedded Class0/I phase with those in the Class II phase \citep{Kenyon1990}.  

However, it is extremely challenging to detect such low-mass planets since they may not be able to generate obvious disc features. The mass of these planets ($\lesssim$10 $M_{\oplus}$) is lower than the disc's thermal mass (especially at the outer disc), 
\begin{equation}
M_{th}=\frac{c_s^3}{G\Omega_p}\approx 28 M_{\oplus}\left(\frac{c_s}{0.6 {\rm km s}^{-1}}\right)\left(\frac{M_{\odot}}{M_*}\right)^{1/2}\left(\frac{a_p}{5 {\rm au}}\right)^{3/2}\,,
\end{equation}
where $a_p$ is the planet's distance to the central star. Thus, they may not generate observable disc features.

On the other hand, an embedded planet is still attached to the background disc and its gravity leads to density enhancement around the planet, forming an envelope. The planetary envelope with a higher density can extend all the way up to the planet's Hill radius \footnote{  In reality, the density changes smoothly from the disc to the core. Depending on if we compare the planet's gravitational force with the star's tidal force or with the disc's thermal pressure, we can say that the planet's influence is to the Hill or Bondi radius.
For low-mass planets studied in this paper, the Hill radius will be larger than the Bondi radius so that we use the Hill radius as the upper limit on the planet's influenced region.}.
If we can probe into the planet's Hill sphere, we may be able to detect this denser envelope. In this paper,
we show that radio observations can probe into this region and directly reveal the envelopes around low-mass planets. We consider detecting the planet's envelope as detecting the planet itself.  The methods
are introduced in Section \ref{sec:method}, including both 1-D and 3-D models. The results are presented in Section \ref{sec:results}. Then, we apply our model to the clump
in TW Hydrae disc in Section \ref{sec:twhydrae}.
After a short discussion in Section \ref{sec:discussion}, the paper is concluded
in Section \ref{sec:conclusion}.

\section{Methods }
\label{sec:method}
We calculate the envelope structure using both 1-D models and 3-D simulations. The 1-D models
further allow us to calculate the envelope's evolution over millions of years, while 3-D simulations calculate the detailed envelope structure
around the planetary core and provide more accurate thermal emission. 

\subsection{The 1-D Model for the Envelope Evolution}
\label{sec:1denvevo}
\begin{figure}
\includegraphics[trim=4mm 8.mm 0mm 1mm, clip, width=3.4 in]{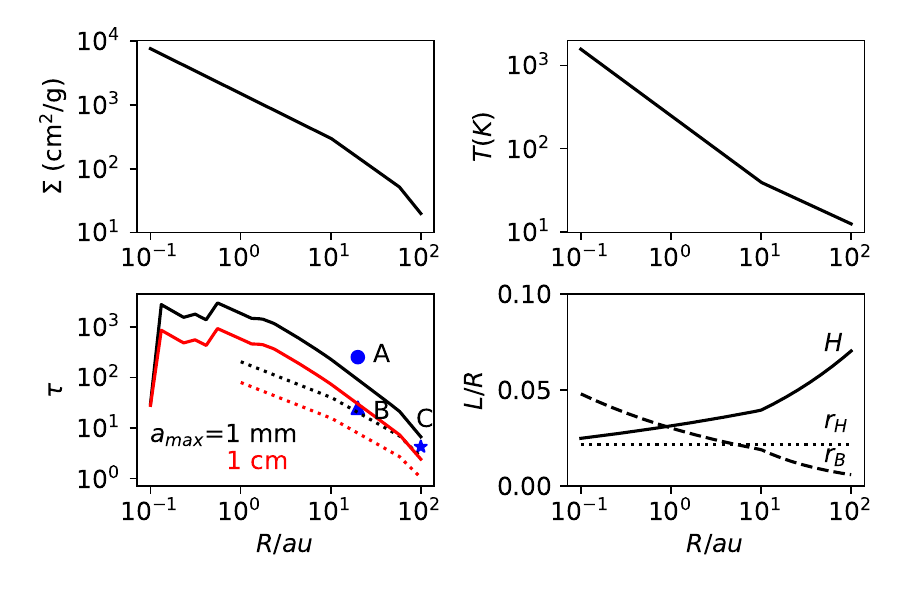}\\
\caption{The radial profiles of the adopted protoplanetary disc structure. The four panels show the surface density, midplane temperature, optical depth, and typical length scales. The solid curves in the optical depth panel show the optical depth using the Rosseland mean opacity, while the dotted curves show the optical depth at 1.3 mm using the monochromatic opacity. The black and red curves in the optical depth panel show the optical depth of the disc with $a_{max}=1$ mm and $a_{max}=1$ cm dust respectively. The solid, dotted, and dashed curves in the length scale panel represent the disc scale height, Hill radius, and Bondi radius of a  10 $M_{\oplus}$ planet at different radii in the disc. The points labeled as A, B, C represent the optical depths (using the Rosseland mean opacity) of Case A, B, C in Section \ref{sec:3demission}. }
\label{fig:radialstr}
\end{figure}

To calculate the structure of the planetary envelope, we first need to specify a protoplanetary disc structure which serves as the outer boundary conditions for the planetary envelope calculation.  
We adopt the disc structure used in \cite{Zhu2021}, which fits the structure from the disc thermal calculation of \cite{Alessio1998}. The disc midplane temperature is
\begin{equation}
T_{mid}=\begin{cases}
39.4\times (R/10\, au)^{-4/5}\; {\rm K} \;\;\;\; if\;\;\;R<10\;{\rm au}\\
39.4  \times (R/10\, au)^{-1/2}\; {\rm K} \;\;\;\; if\;\;\;R>10\;{\rm au}\,,\label{eq:Tc}\\
\end{cases}
\end{equation}
where the stellar irradiation dominates in the disc region beyond 10 au. In this work, we use $R$ to present the cylindrical distance to the star in the disc, and $r$ to present the spherical distance to the planetary core within the planetary envelope. 
We derive the disc surface density from the $\alpha$ disc model \citep{Shakura1973} with 
the additional requirement that the disc needs to be gravitationally stable.
Thus, we have surface density $\Sigma=min\{\dot{M}/(3\pi\nu),\Sigma_{Q=1}\}$, where $\nu=\alpha c_{s}^2/\Omega$
and $\Sigma_{Q=1}=c_{s}\Omega/(\pi G)$. We choose a constant accretion rate throughout the disc $\dot{M}=10^{-8}M_{\odot}/yr$ with $\alpha=10^{-3}$. The disc's radial structure is shown in Figure \ref{fig:radialstr}.
At the disc radii $R$=20, 50, and 100 au, the disc's surface densities are 151, 60, and 20 g/cm$^2$, the midplane temperatures ($T_{mid}$) are 27.8, 17.6, and 12.4 K, the
disc's aspect ratios (H/R) are 0.047, 0.059, and 0.07, and the midplane densities ($\rho_{mid}$) are 4.3$\times$10$^{-12}$, 5.4$\times$10$^{-13}$, and 7.5$\times$10$^{-14}$ g cm$^{-3}$.

To derive the envelope structure, we adopt the opacity compiled in \cite{Zhu2021}, including the dust opacity from \cite{Birnstiel2018}, the molecular opacity from \cite{Freedman2014},
and the atomic opacity from \cite{Colgan2016}.  Dust opacity is the main opacity source in the disc beyond 0.1 au where $T\lesssim$1500 K .
Different opacities due to 
different dust size distributions can significantly affect the planet evolution \citep{Pollack1996}. Thus, we calculate dust opacities with different
dust size distributions to explore their effects. 
These dust distributions follow the nominal $q=3.5$ power law with a minimum particle size of 10$^{-5}$ cm, but  having different maximum particle sizes. 
The Rosseland mean opacity at different disc radii and the opacity at different wavelengths are listed in Table \ref{tab:opacity} \footnote{Scattering is non-negligible for dust opacity. For the mean opacities, the Rosseland mean uses the total opacity/extinction while the Planck mean uses the absorption opacity only. For the monochromatic opacity, the absorption opacity can be calculated with $\kappa_{\nu,tot}(1-\omega_{\nu})$, where $\kappa_{\nu,tot}$ is the total opacity and $\omega_{\nu}$ is the scattering albedo.}. The disc's optical
depths  using the Rosseland mean opacity and the monochromatic opacity at 1.3 mm are shown in Figure \ref{fig:radialstr}. {  Since the disc's optical depth using the Rosseland mean opacity is greater than 1 within 100 au, we use the radiative diffusion approximation to calculate the disc's thermal structure. This assumption breaks down if planets are forming in the dust-depleted disc regions, such as within a disc cavity or in a gas-poor disc during the late stages of disc evolution. In those cases, the envelope structure needs to be calculated using the more detailed two-stream approximation \citep{Lee2018}.}

\begin{table}
\centering
\caption{Dust opacity at different disc radii and wavelengths}
\begin{tabular}{ |c|c|c| } 
 \hline
 & a$_{max}$=1 mm & a$_{max}$=1 cm \\
   \hline
   \multicolumn{3}{l}{Rosseland Mean $\kappa$}\\
   \hline
20au (27.8K) & 0.60 cm$^{2}/g$ & 0.20 cm$^{2}$/g\\ 
50au  (17.6K) & 0.43 cm$^{2}$/g & 0.15 cm$^{2}$/g\\ 
100au (12.4K) & 0.34 cm$^{2}$/g & 0.12 cm$^{2}$/g \\ 
\hline
 \multicolumn{3}{l}{dust ($\kappa_{\nu,tot}$, $\omega_{\nu}$)$^a$}\\
 \hline
0.85 mm &(0.18 cm$^{2}$/g, 0.80) &(0.066 cm$^{2}$/g, 0.76) \\
1.3 mm &(0.13 cm$^{2}$/g, 0.86) &(0.052 cm$^{2}$/g, 0.81) \\
3 mm &(0.058 cm$^{2}$/g, 0.95) &(0.033 cm$^{2}$/g, 0.89)\\
1 cm &(0.001 cm$^{2}$/g, 0.89) &(0.016 cm$^{2}$/g, 0.96)\\
\hline
 \hline
\end{tabular}
$^a$The total opacity and scattering albedo at different wavelengths.
\label{tab:opacity}
\end{table}

With the given disc structure, we calculate the envelope structure of the planets at 20, 50, and 100 au, using the 1-D model in \cite{Zhu2021}.
Using our adopted Rosseland mean opacity ($\kappa(r)=\kappa(\rho(r),T(r))$) and the Equation of State (EoS, $\rho(r)=\rho(T(r),P(r))$) from \cite{Piso2015}, we solve the structure equations
\begin{align}
\frac{dP(r)}{dr}&=-\frac{Gm(r)}{r^2}\rho(r)-\frac{GM_{*}r}{a_{p}^3}\rho(r)\label{eq:Pr}\\
\frac{dT(r)}{dr}&=\nabla(r)\frac{ T(r)}{P(r)}\frac{dP(r)}{dr}\\
\frac{dm(r)}{dr}&=4\pi r^2 \rho(r)\label{eq:mr}\\
\frac{dL(r)}{dt}&=4\pi r^2 \rho(r)\left(\epsilon-T(r)\frac{dS(r)}{dt}\right)\label{eq:Lr}\,,
\end{align}
where $r$ is the radial distance to the planet center, $a_p$ is the planet's distance from the central star, 
$P(r)$, $T(r)$, and $L(r)$ are the radial profiles
of the pressure, temperature, and luminosity, $m(r)$ is the mass enclosed within the radius $r$, and $\epsilon$
is the rate of the internal energy generation without considering the atmosphere contraction.  
Dissipative drag due to planetesimal accretion and pebble accretion all
contributes to $\epsilon$. The term with $\frac{dS(r)}{dt}$ is the energy input from the envelope's
gravitational contraction, and determines the envelope's time evolution. Following \cite{Piso2014}, we use
global energy balance to calculate the envelope contraction and assume $\epsilon$=0. Thus, $L(r)$ is a constant
in our model.

The temperature gradient
$\nabla(r)\equiv d {\rm ln} T(r)/d{\rm ln} P(r)$ depends on the radial energy transport process. For a radiative envelope,
it is determined by the radiative diffusion. On the other hand, when
the envelope becomes convective, we have $\nabla(r)=\nabla_{ad}$
due to the  efficient energy-transport by convection (as simulated in \citealt{Zhu2021}). Overall, we have
\begin{equation}
\nabla(r)=min\bigg\{\frac{3\kappa(r)P(r)}{64\pi G m(r)\sigma T(r)^4}L(r),\nabla_{ad}\bigg\}\,,
\end{equation}
where $\sigma$ is the Stefan-Boltzmann constant, and the adiabatic temperature gradient $\nabla_{ad}$ is calculated using the realistic EoS in
\cite{Piso2015}.

Using the disc density and temperature as the outer boundary conditions for the Equations \ref{eq:Pr}-\ref{eq:mr}, we can adjust the envelope luminosity to search for the static envelope solution
having the mass of $m_{env}$.  More specifically, we integrate the structure equations from the local disc scale height ($H$) inwards to the core radius ($(m_{core}/10 M_{\oplus})^{1/3} \times 10^{-4}$ au) using the Runge-Kutta method. 
Our approach of using the disc condition as the local disc scale height as the outer boundary is different from previous 1-D models which use the minimum of the planet's Hill radius ($r_{Hill}$) and Bondi radius ($r_{Bondi}$) as the outer boundary (e.g. \citealt{Bodenheimer1986}). 
Our choice together with the $GM_{*}r\rho(r)/a_{p}^3$ term in Equation  \ref{eq:Pr} allows the envelope to transit smoothly to the background disc density \citep{Zhu2021}. 
At the disc scale height, we set the density as 
\begin{equation}
\rho(H)=1.33 \rho_{mid}e^{-1/2}=0.81 \rho_{mid}\,.\label{eq:Hmid}
\end{equation}
In a 3-D disc, the density structure is not spherically symmetric around the planet. The factor of 1.33 in Equation \ref{eq:Hmid} considers this effect at $H$ away from the planet center.
Considering that the envelope is normally radiative at $H$, the temperature at $r=H$ can be derived using
\begin{equation}
T(r)^4=\frac{3 L}{4\pi c a_r}\int_{r}^{\infty}\frac{\kappa\rho_{mid}e^{-r'^2/2H^2}}{r'^2}dr'+T_{mid}^4 \label{eq:Trtest}\,,
\end{equation}
with $r=H$ \citep{Zhu2021}, where $a_r$ in Equation \ref{eq:Trtest} is the radiation constant. 
Then, we can derive $P(H)=\mathcal{R}\rho(H)T(H)/\mu$, where $\mathcal{R}$ is the gas constant  and the mean molecular weight $\mu$ is assume to be 2.35 at $r=H$.
With $P(H)$ and $T(H)$ known, we can integrate the structure equations by giving a luminosity $L$. We iterate the $L$ values until
the envelope mass within the planetary boundary $r_p$ equals the desired mass $m_{env}$. Based on the bounded region revealed in simulations of \cite{Zhu2021}, 
$r_p$ is chosen as 1/10 of $min\{r_{B}, r_{H}\}$. We have verified that using $r_p=min\{r_{B}, r_{H}\}$ does not affect our results.

After varying the envelope mass $m_{env,i}$, we derive the structure of various envelopes and record their corresponding luminosities
$L_{i}$. Then, we can use the global energy balance to derive the evolutionary time between the static solutions with $m_{env,i}$ and $m_{env,i+1}$,
\begin{equation}
\Delta t=\frac{-E_{i+1}+E_{i}}{(L_{i}+L_{i+1})/2}\,, \label{eq:deltat}
\end{equation}
where $E_{i}$ is the addition of the gravitational and internal energy with the envelope mass of $m_{env,i}$.
\footnote{ As shown in Figure 5 of \cite{Piso2014}, the next order correction beyond Equation \ref{eq:deltat} is constructing a model where luminosity is calculated self-consistently using entropy difference at each layer between successive solutions. The effects of mass and volume change on the energy balance add another smaller corrections. More sophisticated models are needed in future to account for these effects.}
Finally, connecting all the envelopes
from small to large masses, we can derive the evolution of the envelope during its KH contraction. 

\begin{figure*}
\includegraphics[trim=5mm 6mm 6mm 6mm, clip, width=3.4in]{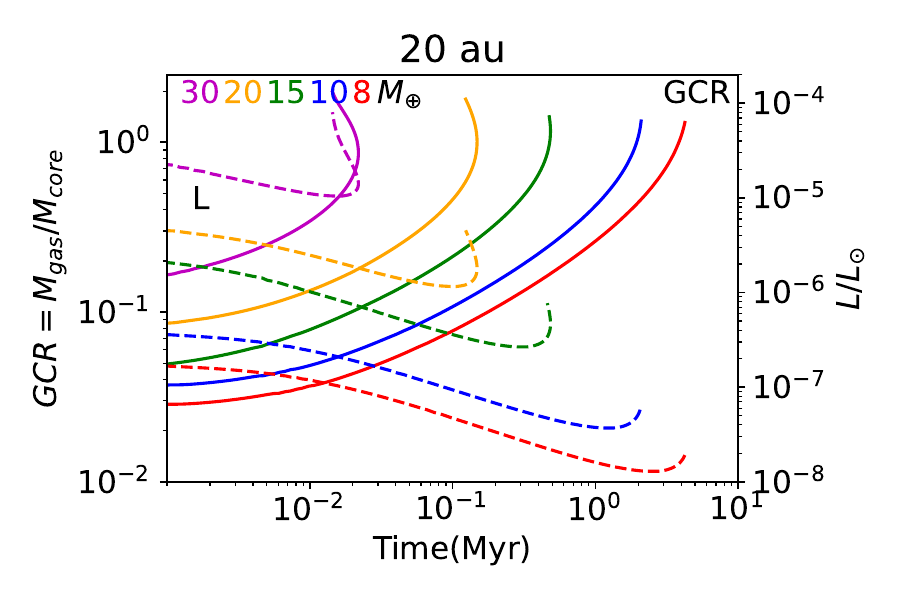}
\includegraphics[trim=5mm 6mm 6mm 6mm, clip, width=3.4in]{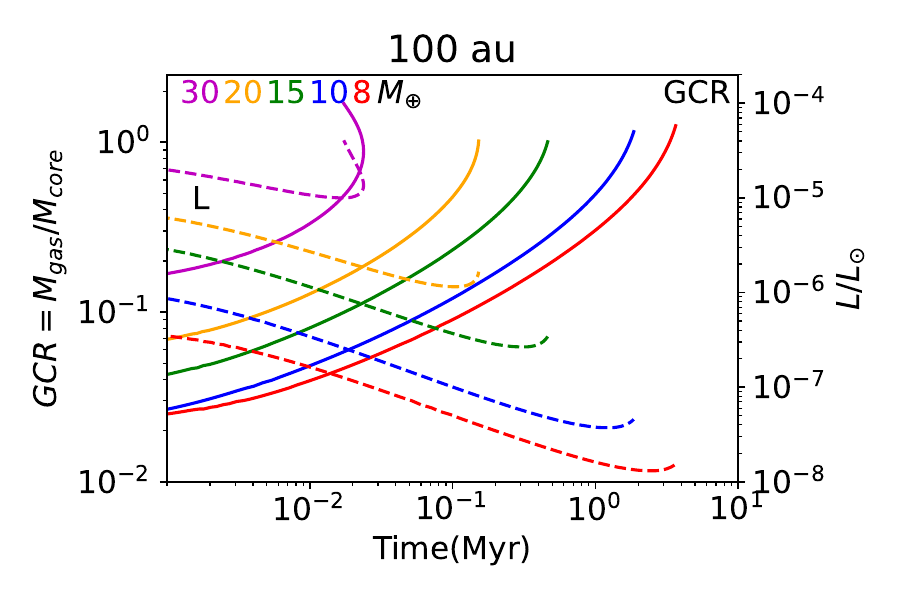}\\
\caption{The evolution of the gas-to-core mass ratio (solid curves) and the luminosity (dashed curves) for different planetary core masses (different colors) at different
disc radii (20 au on the left, and 100 au on the right). The dust opacity is calculated with the $a_{max}$=1 mm dust population. }
\label{fig:evolution}
\end{figure*}

\subsection{The 1-D Model for the Envelope's Thermal Emission }
\label{sec:1demission}
After deriving the envelope structure, we can calculate the thermal emission from these envelopes. We choose six characteristic luminosities ( $L$=$1.55\times10^{-7}$, $4.64\times10^{-7}$, $1.55\times10^{-6}$, $4.64\times10^{-6}$, $1.55\times10^{-5}$, and $4.64\times10^{-5}$ $L_{\odot}$) to represent
different evolutionary stages during the KH contraction. These luminosities correspond to $L=G\cdot 10 M_{\oplus}\cdot\dot{M}/2R_{\oplus}$, with $\dot{M}=10^{-6}$, $3\times10^{-6}$, $10^{-5}$, $3\times10^{-5}$, $10^{-4}$, and $3\times10^{-4}$ $M_{\oplus}/yr$. The lowest accretion rate is equivalent to gathering
a 10$M_{\oplus}$ envelope within the disc's lifetime, $\sim$ 10 million years. It is the lowest accretion rate the planet should have in order to reach run-away
accretion before the disc dissipates. 
Overall, these luminosities cover the expected luminosity range during the planet's KH contraction/envelope gathering phase, as in Figure \ref{fig:evolution}. 

With the given luminosity and the core mass, we first derive the envelope structure using our 1-D model in Section \ref{sec:1denvevo}.
Then, we generate an axisymmetric 2-D cylindrical grid in the disc's radial and vertical directions to cover the region
around the planet.  Here, we assume that the density and temperature structures of the envelope and the background disc
are axisymmetric with respect to the disc's rotational axis (z axis) to simplify the computation. In reality, non-axisymmetric spiral arms will be generated, which will be discussed with our 3-D simulations in Section \ref{sec:3dmodels}.   
We have 700 grid cells uniformly spaced from 0 to 2.5$H$ in the radial direction, and 8000 grid cells uniformly spaced from -3$H$ to 3$H$ in the vertical direction. We choose more grid cells in the vertical direction since the disc is vertically stratified. The results hardly change even with a much coarser resolution.
For the grid cell whose distance to the planet center is larger than $H$, its density is assigned following the background Gaussian disc density profile  $\rho_{mid} exp\{-z^2/2H^2\}$.
For the grid cell within $H$ from the planet center, we interpolate the 1-D planet envelope structure to derive the density. The temperature of each grid cell is assigned in the similar fashion, except that we allow a smoother transition. When the grid cell is further away from the planet center by 2$H$, its temperature is assigned with
the uniform background disc temperature. When it is between $H$ and 2$H$ from the planet center, its temperature is calculated using Equation \ref{eq:Trtest} where $r$ is replaced with the grid cell's $r$. When the grid cell is within $H$ from the planet center, we interpolate the temperature from the 1-D planet envelope structure. Overall,
the above procedure allows us to embed a spherical envelope into a cylindrical disc region. 

When the density and temperature structures of the grid cells are assigned, we use the ray-tracing method to calculate the intensity coming out of the disc surface,
\begin{equation}
I(\bar{r})=\int_{-3H}^{3H} S_{\nu}\rho \kappa_{\nu,tot} e^{-\tau_{\nu}} dz\,,
\end{equation}
where $\bar{r}$ is the horizontal distance to the planet center at the disc midplane. We have
\begin{equation}
S_{\nu}=\omega_{\nu}J_{\nu}(\tau_{\nu})+(1-\omega_{\nu})B_{\nu}\,,
\end{equation}
with
\begin{align}
&J_{\nu}(\tau_{\nu})=B_{\nu}(T)\nonumber\\
&\times\left(1+\frac{e^{-\sqrt{3(1-\omega_{\nu})}\tau_{\nu}}+e^{\sqrt{3(1-\omega_{\nu})}(\tau_{\nu}-\tau_{\nu,t})}}{e^{-\sqrt{3(1-\omega_{\nu})}\tau_{\nu,t}}(\sqrt{1-\omega_{\nu}}-1)-(\sqrt{1-\omega_{\nu}}+1)}\right)\,, \label{eq:Jnu}
\end{align}
where $\tau_{\nu,t}(\bar{r})$ and $\tau_{\nu}(\bar{r},z)$ are the total and variable optical depth along the vertical direction. Although Equation \ref{eq:Jnu} is derived based on
the plane parallel assumption \citep{Miyake1993,Zhu2019}, we still apply it to each radial position considering that the radiation field is dominated by the disc's 
plane parallel radiation. Then, we integrate the intensity emerging from the disc to derive the excess flux emitted by the planetary envelope 
from an area $\pi H^2$ around the planet
\begin{equation}
Flux_p=\int_{0}^{H} \left(I(\bar{r})-I_0(\bar{r})\right)\frac{ 2\pi \bar{r}}{d^2}d\bar{r}\,,\label{eq:Fluxint}
\end{equation}
where $d$ is the distance between the protoplanetary disc and Earth, and we assume $d=100 {\rm pc}$ except for the TW Hydrae calculations.
$I_0(\bar{r})$ is the intensity from a reference calculation with an extremely low-mass planetary core (0.01 $M_{\oplus}$), which represents the background disc emission with our disc and grid setup. Thus, $Flux_p$ captures the excess mass and temperature in the disc due to the introduction of the planet.
The background disc flux is thus
\begin{equation}
Flux_{bg}=\int_{0}^{H} I_0(\bar{r})\frac{ 2\pi \bar{r}}{d^2}d\bar{r}\,.\label{eq:Fluxintbg}
\end{equation}

\subsection{The 3-D Model for the Envelope's Thermal Emission }
\label{sec:3demission}
We also post-process the 3-D radiation hydrodynamical simulations of \cite{Bailey2022} to generate synthetic images. One additional simulation is carried out for studying the clump in TW Hydrae disc. These are 3-D shearing box simulations with a low-mass planet at the center. Mesh-refinement with two refinement levels has been adopted with the highest resolution of 128 cells per disc scale height.
The simulation domain covers -2$H$ to 2$H$ in both the radial and azimuthal directions along the disc plane (the $x$ and $y$ directions in the shearing box setup), and 0 to 2$H$ in the vertical direction. 
The simulations use the dimensionless parameters 
\begin{eqnarray}
\beta&\equiv&\frac{a_r c\left(\frac{\mu m_p}{k}\right)\frac{T_0^3}{\rho_0}}{c_s}\\
\bar{\kappa}&\equiv&\kappa_0\rho_0 H_0\\
q_{t}&\equiv&\frac{M_p a_p^3}{M_* H^3}\,,
\end{eqnarray}
where the subscript 0 refers to the disc midplane quantities.
The physical meaning of $\bar{\kappa}$ and $q_t$ are straightforward, which represent the vertical optical depth and the planet's mass in thermal mass. 
The physical meaning of $\beta$ is more obscure, which represents the ratio between the radiation energy flux and the thermal energy flux. 

All steps to calculate the thermal emission are the same as in Section \ref{sec:1demission}, except that we generate
a 3-D grid and directly import the density and temperature from the 3-D simulation
into the 3-D grid.
We can scale the simulations to our disc conditions. Following
\cite{Bailey2022}, we assume that $\mu$=2 for 3-D simulations. 
We scale the simulation with $\bar{\kappa}=100,\beta=1$, $q_t$=0.5 (denoted as Case A) and the simulation with $\bar{\kappa}=10,\beta=1$,  $q_t$=0.5 (denoted as Case B) to our disc condition at 20 au, and the new simulation with $\bar{\kappa}=1.7,\beta=10.28$,  $q_t$=0.213 (denoted as Case C)
to 100 au. More specifically, the 20 au simulations correspond to $T_0$=27.8 K, $\rho_0=3.48\times10^{-12}$ g cm$^{-3}$,
$\Sigma=133$ g cm$^{-2}$ with a 22 M$_\oplus$ planet. The $\bar{\kappa}$=100 case (Case A) has the total Rosseland mean optical depth of 251, while 
the $\bar{\kappa}$=10 case (Case B) has the Rosseland mean optical depth of 25.1. As shown in the lower left panel of Figure \ref{fig:radialstr}, the optical depth of 251 is close to the Rosseland mean optical depth of our disc at 20 au with $a_{max}=$1 mm particles, and 25.1 is close to that with $a_{max}=$1 cm.
Thus, we use monochromatic opacities calculated with
$a_{max}=1$ mm and 1 cm dust populations to post-process $\bar{\kappa}=100$ (Case A) and 10 (Case B) simulations respectively. The Case C simulation
corresponds to $T_0$=12.4 K, $\rho_0=4.53\times10^{-14}$ g cm$^{-3}$,
$\Sigma=13$ g cm$^{-2}$ with a 32 M$_\oplus$ planet at 100 au, and its total Rosseland mean optical depth is 4.3 with $a_{max}=1$ mm dust. 

We can convert intensity to brightness temperature using
\begin{equation}
T_{B}(\nu)\equiv\frac{I_{\nu}c^2}{2k\nu^2}\,.
\end{equation}
We can also convolve the map of $T_B$ with some Gaussian kernel to derive synthetic observations. The Gaussian kernel has a $\sigma$ which is related to the Full Width at Half Maximum (FWHM) as $\theta$ = 2.355 $\sigma$. {  The observed flux density can be written as
\begin{equation}
S_{\nu}=\frac{2k T_B \nu^2 \Omega_{bm}}{c^2}\,, \label{eq:snu}
\end{equation}
where $\Omega_{bm}$ is the beam's solid angle. With a Gaussian beam, $\Omega_{bm}=\pi \theta_{a}\theta_{b}/[4 \rm{ln}(2)]$ where $\theta_{a}$
and $\theta_{b}$ are the beam major and minor axis. }



\section{Results}
\label{sec:results}

\begin{figure*}
\includegraphics[width=6.8 in]{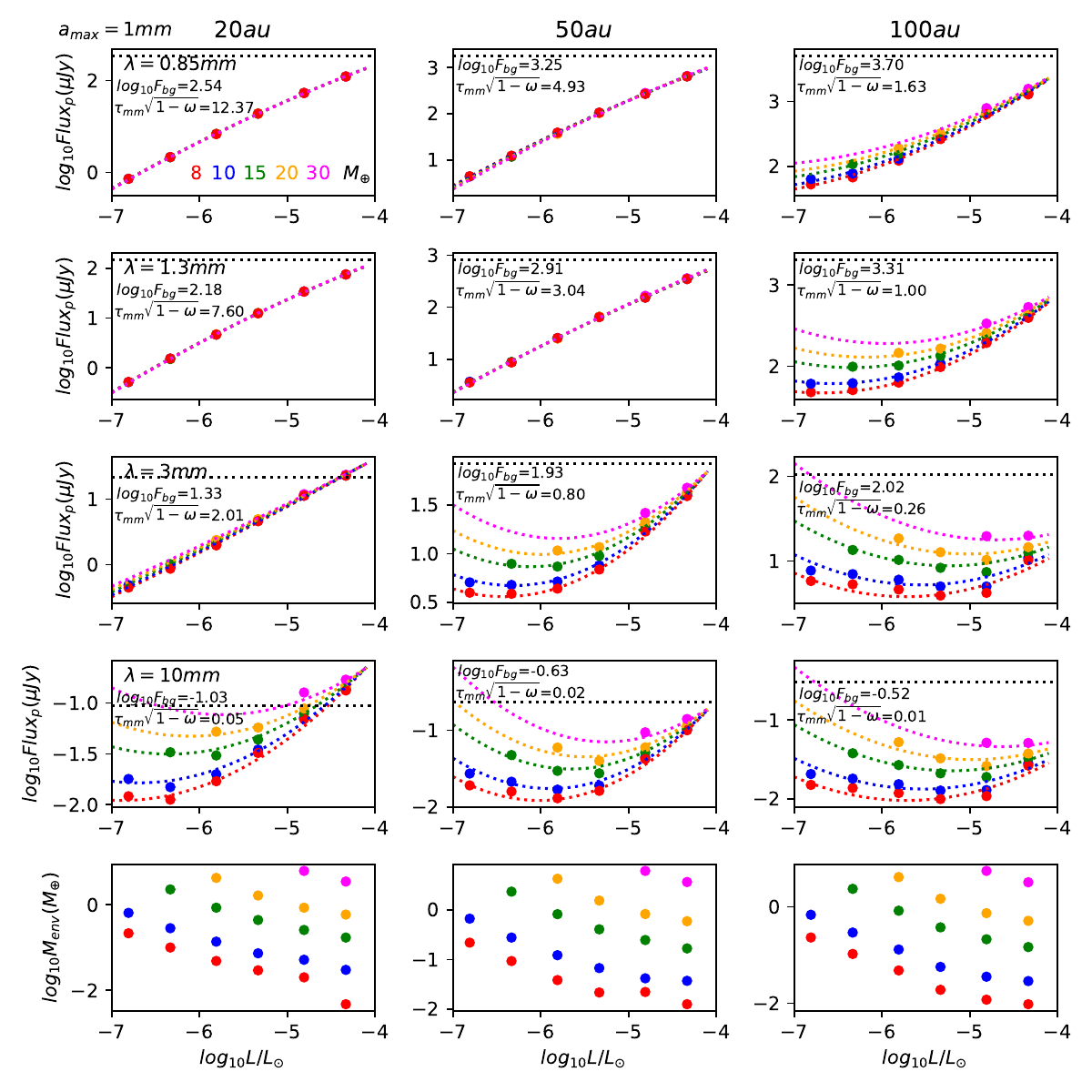}\\
\caption{Upper four rows: the excess thermal flux at 0.85, 1.3, 3, and 10 mm (from the top to bottom rows) from the planetary envelope integrated in the disc plane within 1 disc scale height from the planet center (Equation \ref{eq:Fluxint}).  We adopt the $a_{max}$=1 mm dust opacity. The bottom panels: the mass ratio between the envelope mass within $r_p$ and the core mass. The flux from the background disc (in the unit of $\mu$Jy) within the same region ($\pi H^2$) is plotted as the horizontal dotted line and also labeled in the upper left corner of each panel. The background disc flux increases with the disc radii since the integrated region ($\pi H^2$) increases with the disc radii.
The effective optical depth $\tau_{mm}\sqrt{1-\omega_{mm}}$ is also shown in the upper left corner of each panel. Different colors represent planets with different core masses. The dotted curves are fits to the data points. In the extremely optically thick cases, the dots and curves with different colors overlap. The left to right columns represent the planets at 20, 50, and 100 au. }
\label{fig:1mm}
\end{figure*}

\begin{figure}
\includegraphics[trim=1mm 18mm 1mm 18mm, width=3.3 in]{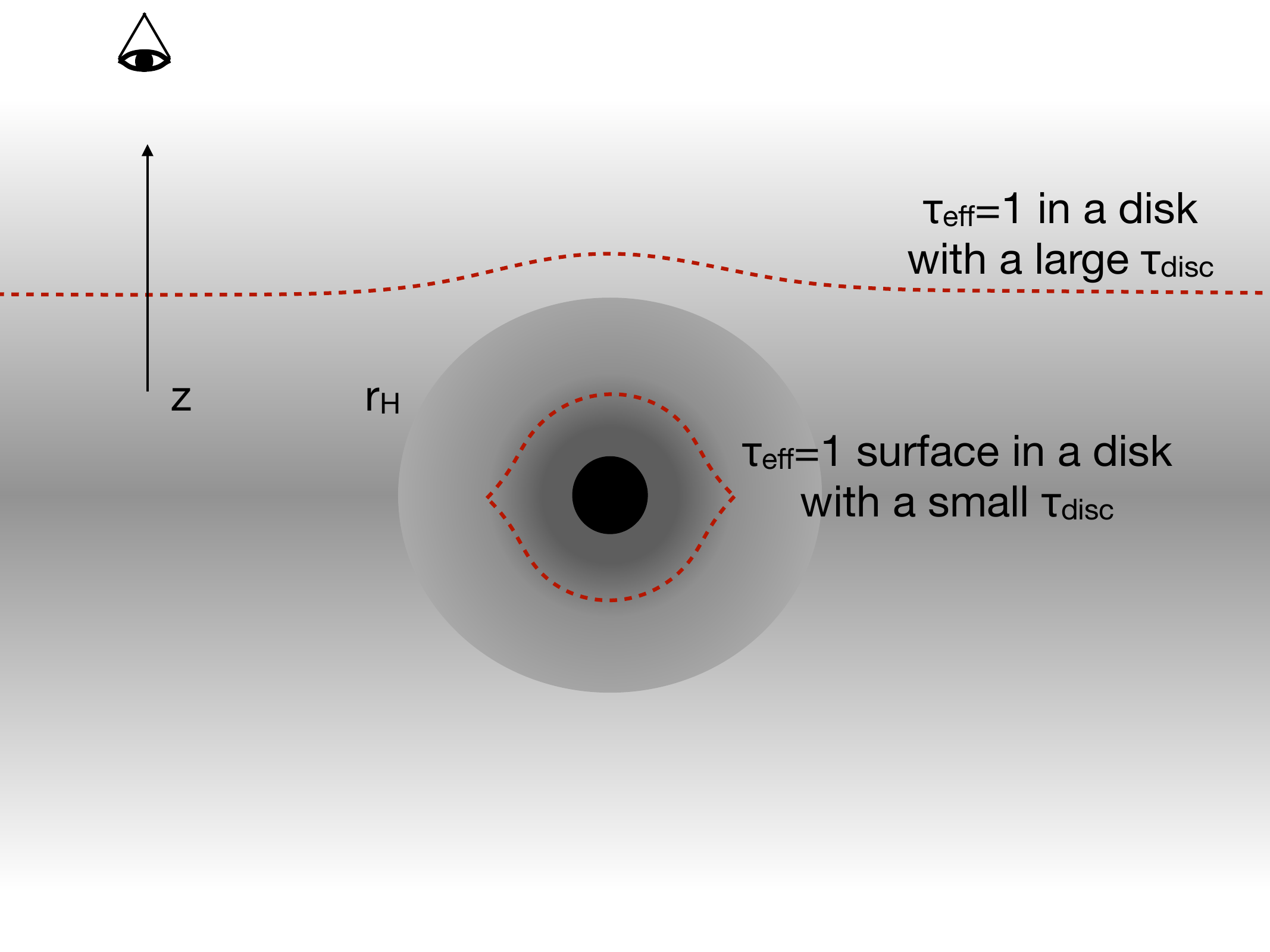}\\
\caption{The schematic plot showing the envelope around the embedded low-mass planet. When the disc's optical depth is small, observations can probe the dense envelope within
the Hill radius, revealing the planetary envelope. }
\label{fig:schematic}
\end{figure}

\subsection{1-D models}
\label{sec:1dmodels}

To show the range of the planet mass and luminosity expected during the KH contraction in the core accretion model, we present
the planet evolutionary tracks at different disc radii in Figure \ref{fig:evolution}. The planet spends most its time with the lowest possible luminosity. 
When the planet's envelope-to-core mass ratio reaches $\sim$1, the envelope starts the run-away accretion.
The luminosity will increase dramatically afterwards, which is not shown in the figure. Thus, we define the run-away time as the time when the gas-to-core mass ratio reaches 1.
Using our nominal opacity with $a_{max}$= 1 mm dust, planetary cores with $\gtrsim$10 M$_{\oplus}$ can undergo run-away accretion within the
disc's lifetime $\sim$3 Myrs. We have verified that, with a lower opacity (e.g. $a_{max}$=1 cm), 
the envelope's KH phase before the run-away accretion is accelerated
and  even lower mass planets can undergo run-away accretion before the disc dissipates. 

We notice that the evolutionary tracks in Figure \ref{fig:evolution} are almost identical no matter where the planet is . This is different from
\cite{Piso2014,AliDib2020} which suggest that the KH contraction is faster at the outer disc and thus
the critical core mass is lower at the outer disc. This difference is due to the different opacity laws adopted.
\cite{Piso2014,AliDib2020} use  ISM dust opacity which increases steeply with temperature. Then, for a planet at the outer cooler disc,
the opacity is relatively low so that the planet can cool faster. Our opacity has a shallow temperature dependence similar to the opacity adopted in \cite{Piso2015} where
the derived critical core mass changes from 8 $M_{\oplus}$ at 20 au to 6 $M_{\oplus}$ at 100 au with the $a_{max}=$1 cm dust opacity.
Since our opacity is slightly higher with $a_{max}$= 1 mm dust, our critical core mass is $\sim$10 $M_{\oplus}$.  Our results are consistent with \cite{Lee2014, Chen2020}.

The higher the core mass is, the higher luminosity the envelope has. Thus, based on the traditional core accretion model,
it should be easier to detect planets with higher mass cores, but they can be short-lived before the run-away accretion.

To calculate the envelopes' thermal radiation at mm-cm bands, we pick some envelopes with different luminosities around different mass cores at different disc radii, and
post-process them with ray-tracing. 
We integrate the intensity in the disc plane over an area within $H$ from the planet center (Equation \ref{eq:Fluxint}). The values of $H$ at different disc radii are given in the lower-right panel of Figure \ref{fig:radialstr}. 

The radio fluxes are shown in Figure \ref{fig:1mm}. More massive cores only have solutions when their luminosities are high, which is consistent with their higher luminosity tracks in Figure \ref{fig:evolution}.
Figure \ref{fig:1mm} also shows that, with the same core mass, the planetary envelope has a stronger radio emission when the planet's luminosity is higher, due to a hotter envelope. The upper left panels in Figure \ref{fig:1mm} show that, when the background disc is optically thick at the observed wavelength, e.g. the inner disc or being observed at shorter wavelengths, the flux of the embedded planet only depends on its luminosity and has no dependence on the planet mass. This is because the envelope density only deviates from the background disc density when the envelope is within the planet's Hill radius (Figure \ref{fig:schematic}). If the disc is already optically thick beyond the planet's Hill radius, the envelope's density at the $\tau_{mm}=1$ photosphere will be quite similar to the disc's density at the $\tau_{mm}=1$ surface no matter how massive the planet is, and thus it is only the excess temperature there due to the planet's luminosity that leads to the excess emission.

More specifically, the Hill radii of our studied planets range from 0.02 $a_{p}$ with $M_p=8\, M_{\oplus}$ to 0.03 $a_{p}$ with $M_p=30\, M_{\oplus}$, while the disc scale height is $\sim$0.051-0.076 $a_p$ from 20 to 100 au. Thus, although the planets are embedded in the disc, their Hill radii are a moderate fraction of the disc thickness. When the disc is optically thick (Figure \ref{fig:schematic}), the region beyond the planet's Hill radius will likely be optically thick. The effective optical depth considering the scattering ($\tau_{eff}\equiv \tau_{mm}\sqrt{1-\omega_{mm}}$ where $\sqrt{1-\omega_{mm}}$ is due to the random walk led by dust scattering, \citealt{Rybicki1979, Zhu2019}) is shown in the upper left corner of each panel {  in Figure \ref{fig:1mm}} \footnote{When the disc is optically thick, the optical depth calculated with the total opacity appears in the energy diffusion equation. When the disc is optically thin, the optical depth with the absorption opacity is used to calculate the emission. Here we define an effective optical depth, which considers the scattering and appears extensively in Equation \ref{eq:Jnu}.}. Clearly, when $\tau_{eff}$ is larger than 1, the planet flux only depends on the luminosity\footnote{In our 1-D model, the density beyond the disc scale height is set as the background disc density. This ignores any density perturbation from the planet beyond the disc scale height. However, this is a good approximation considering that the Hill radii are much smaller than the disc scale height.}. In these cases, it is the temperature excess at the $\tau_{eff}=1$ photosphere determines the envelope's thermal flux. Based on Equation \ref{eq:Trtest}, the excess temperature $\Delta T\equiv T-T_{mid}$ is $\propto L$ when $T\sim T_{mid}$. Thus, the excess thermal emission
should also be proportional to $L$. This trend roughly agrees with upper left panels in Figure \ref{fig:1mm}. On the other hand, in the more realistic cases where the planets' luminosities are at the lower end, the planets' fluxes are 2 orders of magnitude smaller than the background disc flux, making their detection challenging. 

When the background disc is optically thin, we can see through the disc and probe the denser envelope around the planet (Figure \ref{fig:schematic}). At a given luminosity, a more massive core has a more massive envelope (shown in the bottom panels of Figure \ref{fig:1mm}), leading to a stronger thermal emission. On the other hand, the thermal emission with different core masses converges when the luminosity is high. This is because the envelope's thermal emission is more affected by the high envelope temperature in these high luminosity cases. When the luminosity becomes lower, the density plays a more important role in the thermal emission, and we see bigger flux differences for different core masses. Surprisingly, when the luminosity gets very low, the thermal emission for any given core mass seems to flatten out or even increases with the decreasing luminosity. This is due to the counterbalancing effect between the envelope mass and envelope temperature. When the luminosity decreases, the envelope temperature  decreases but the envelope mass increases (the bottom panels of Figure \ref{fig:1mm}).  Eventually, the envelope temperature is close to the disc temperature, and the thermal emission is only affected by the envelope mass (more discussions in Section \ref{sec:simplemodel}). We want to note that, even mm/cm observations cannot probe very deep in the envelope since the envelope will become optically thick quickly. However, as long as the observations can probe inside the Hill radius of the planet, we can detect the enhanced density within the Hill radius which leads to the additional thermal radiation (Figure \ref{fig:schematic}). {  We can also understand the flattening of the envelope's thermal flux using the $\tau_{mm}=1$ surface argument. When the planet's luminosity decreases, the envelope becomes more massive and the $\tau_{mm}=1$ surface moves higher towards the disc surface. At the same time,
the temperature at the $\tau_{mm}=1$ surface is lower with the lower luminosity. As the luminosity gets lower, the envelope's lower temperature is compensated by the larger emitting area so that the total flux is almost a constant. }

\begin{figure*}
\includegraphics[width=6.8 in]{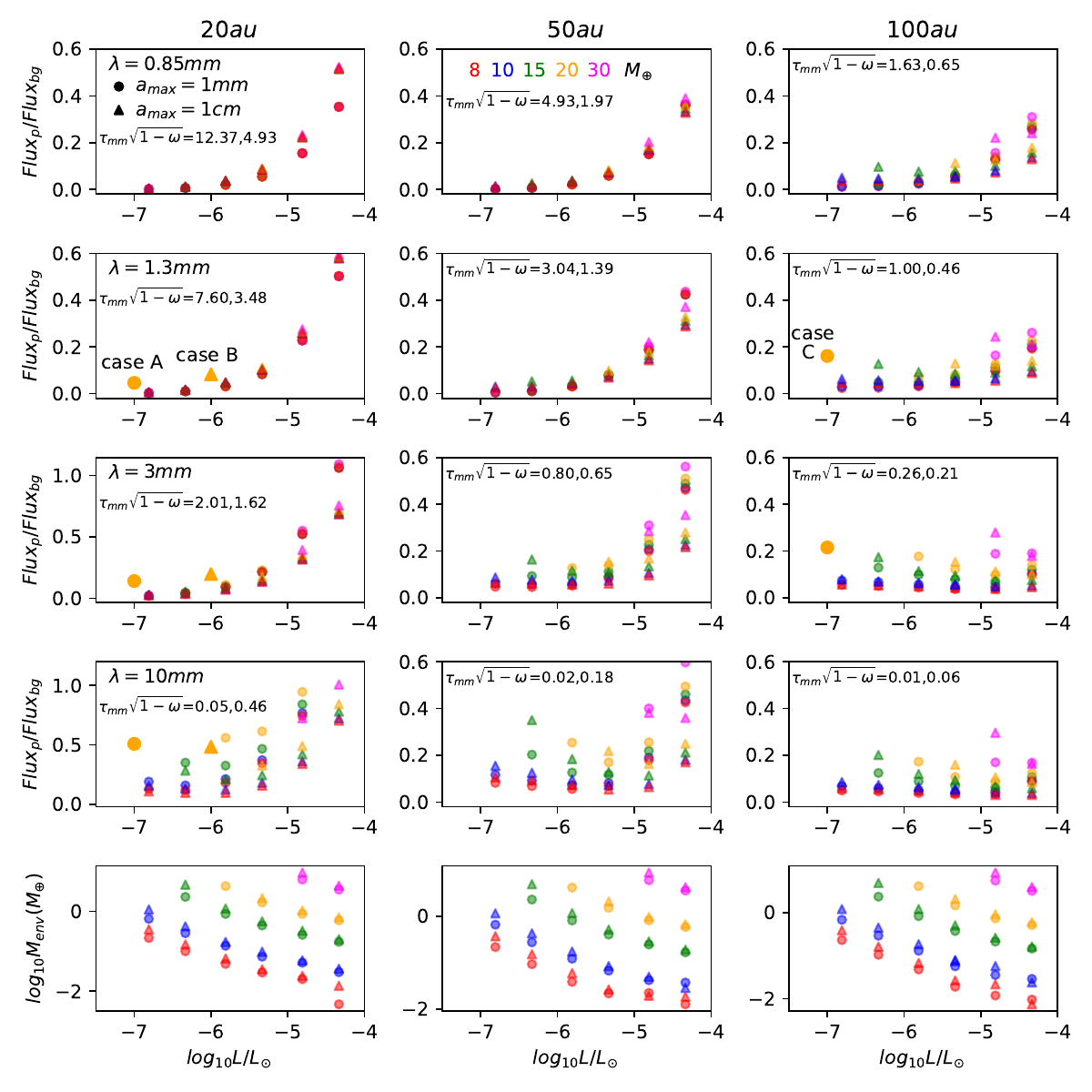}\\
\caption{The ratio between the planet's excess flux and the disc's background flux at different wavelengths (the shortest to longest wavelength from the top to bottom rows) and disc radii (the inner to outer disc from the left to right columns).  The fluxes are integrated over the area within 1 disc scale height from the planet center. Fluxes using two different dust opacity are provided (dots with the $a_{max}$=1 mm dust opacity; triangles with the $a_{max}$=1 cm dust opacity). Planets with different core masses are labeled with different colors. The bottom panels show the planet's envelope mass for each case. The big circles and triangles represent the flux ratios of cases from 3-D simulations. }
\label{fig:fluxrat}
\end{figure*}

On the other hand, these planets are embedded in the disc which itself emits thermal radiation. The  density of the outer envelope scales with the background disc density. It is the flux ratio between the planet's thermal radiation and the background disc radiation that determines if the planet can stand out from the background disc to be detected. The background flux ($Flux_{bg}$) is plotted as the horizontal line and its value is provided in the upper left corner of each panel {  in Figure \ref{fig:1mm}}. For most cases, the planet's excess thermal emission is smaller than the background disc emission. Thus, high sensitivity observations are required to probe the planet's excess emission. The ratio between the planet's excess thermal emission and disc's thermal emission is better shown in Figure \ref{fig:fluxrat}. The flux ratios that are calculated using two different disc opacities are both shown. As expected, when the disc is optically thick, the flux ratio can be very small if the planet luminosity is low. In these cases,
the planets are hidden inside the disc. When the disc is optically thin, we find that the planet's excess emission is higher with a higher mass core. This is because a higher core mass leads to a higher envelope mass within the Hill radius, which leads to a larger flux ratio. We also notice that the flux ratio is generally smaller when the planet is at a larger distance from the star. {  This is mainly because we assume $\pi H^2$ as the beam size to calculate the flux ratio. The outer disc has a larger beam size ($H\propto R^{1.25}$), which results in a stronger background flux within the beam (dotted lines in Figure \ref{fig:1mm}) and thus a smaller flux ratio. } With our adopted beam size, the planet's excess emission is above 10\% of the background emission for most cases we studied, even if the luminosity is very low.  Observations with a higher spatial resolution can  significantly boost this flux ratio since most of the planet's emission is concentrated around the planet while the background emission is more uniform. {  Thus, if the beam size of a real observation is only half of what we assumed and the planet's thermal emission is still mostly within the beam, the resulting flux ratio will be $\sim$4 times higher than those shown in Figure \ref{fig:fluxrat}. On the other hand, a higher resolution in real observations also leads to a smaller flux per beam for the background disc, leading to a worse S/N ratio for the disc and not necessarily a more robust envelope detection. }

\subsection{3-D models}
\label{sec:3dmodels}
\begin{figure*}
\includegraphics[trim=0mm 25mm 0mm 60mm, clip, width=6 in]{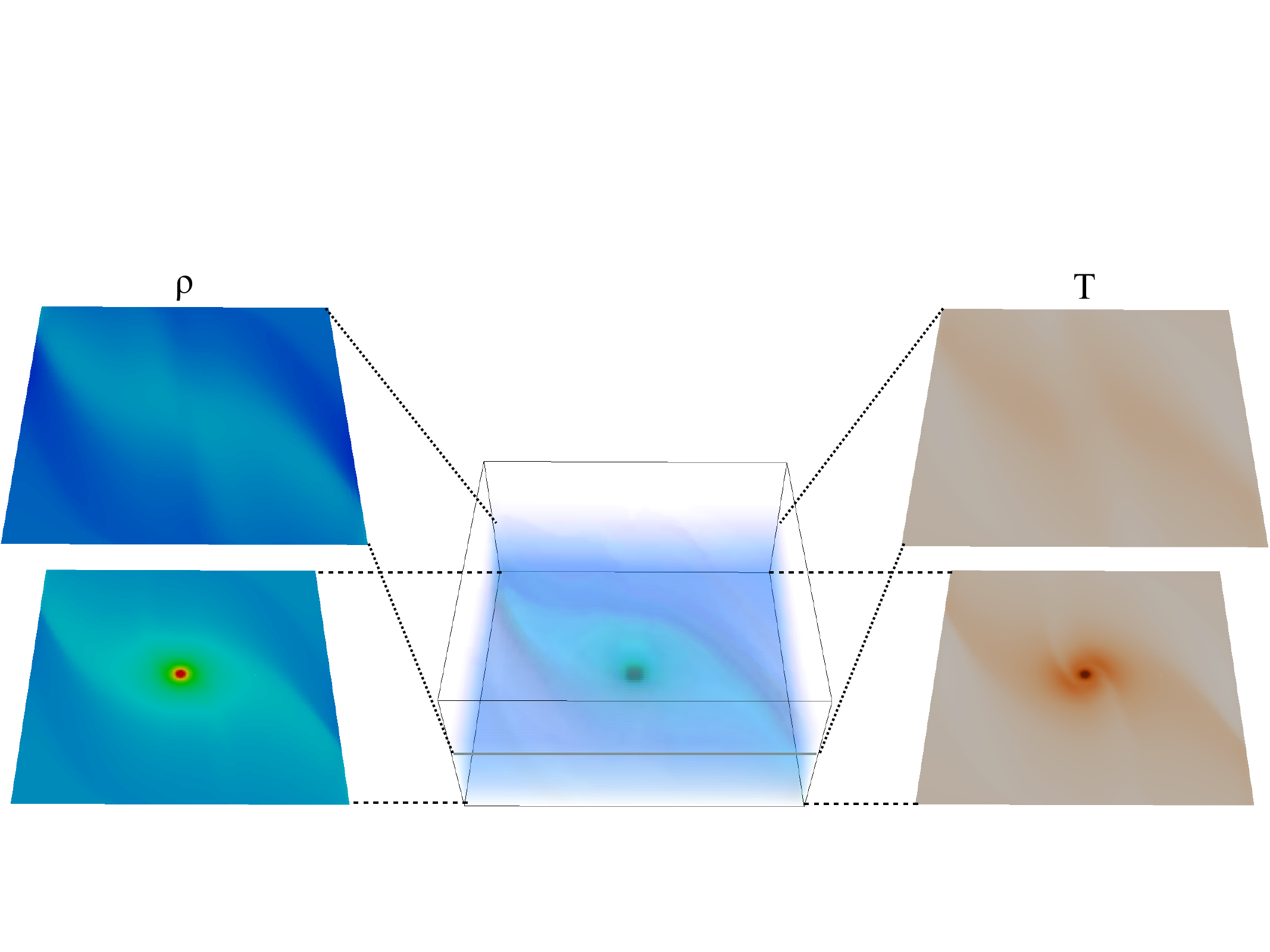}\\
\caption{{  The density (leftmost panels) and temperature (rightmost panels) at the disc midplane (lower panels) and one disc scale height (upper panels) for the 3-D simulation Case C.}  }
\label{fig:3Dsim}
\end{figure*}

\begin{figure*}
\includegraphics[trim=1mm 5.5mm 1mm 6mm, clip, width=5.5 in]{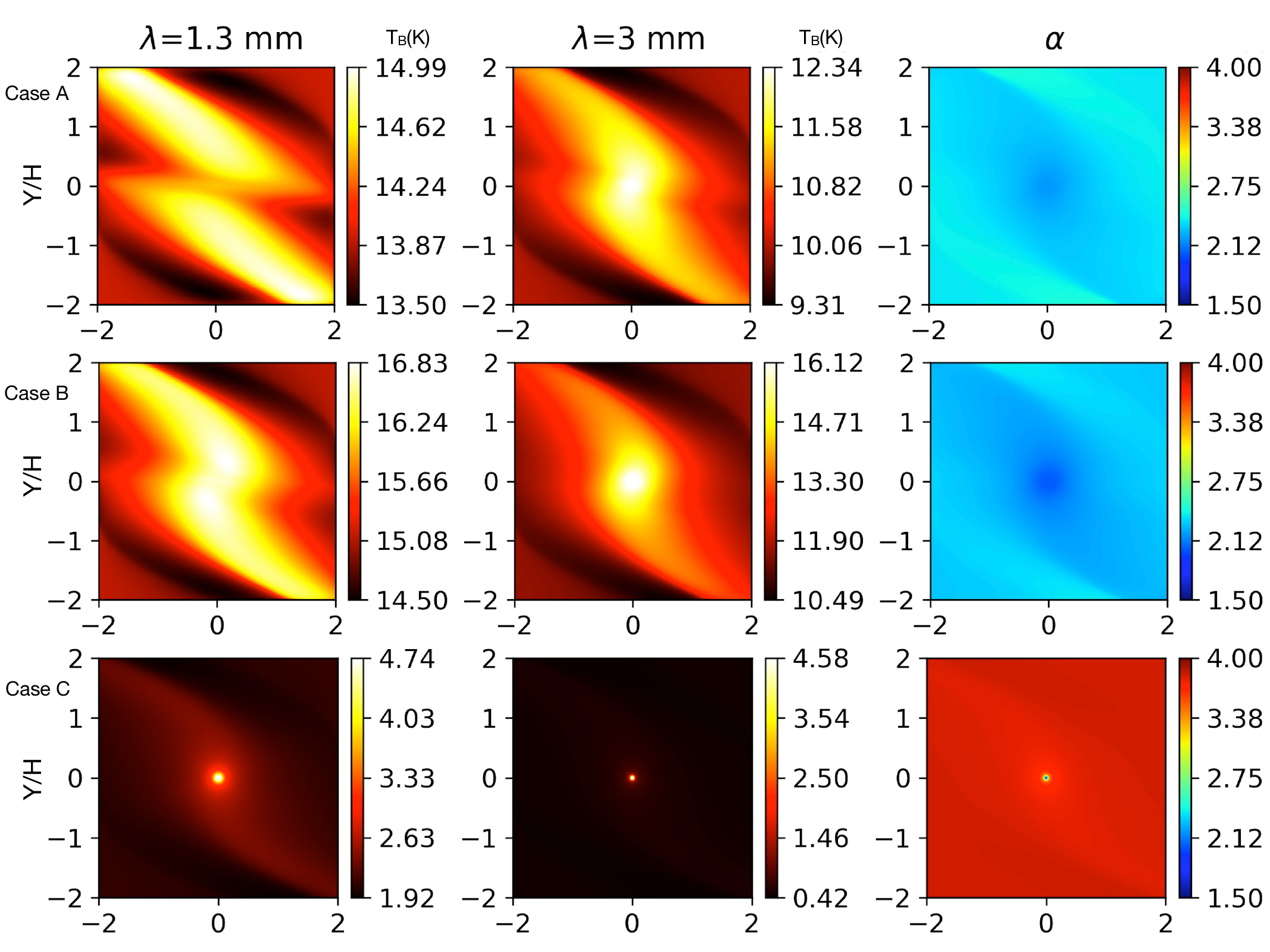}\\
\caption{The synthetic images at two different wavelengths (the left two columns) for the three simulations (the optical depth decreases from the top to bottom panels). The rightmost column shows
the spectral index between these two wavelengths. }
\label{fig:syn}
\end{figure*}

\begin{figure*}
\includegraphics[trim=1mm 5.mm 1mm 7mm, clip, width=5.5 in]{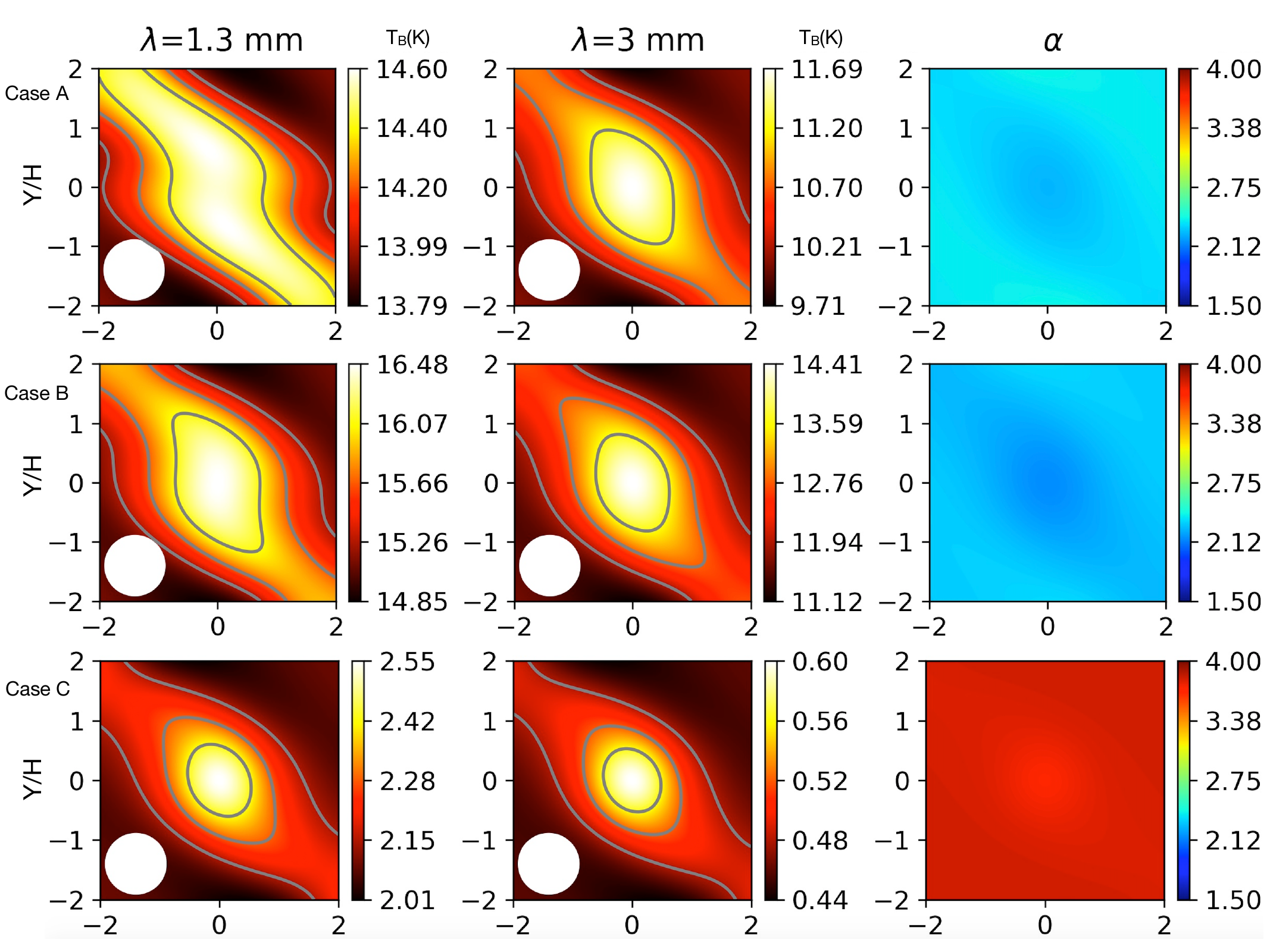}\\
\caption{Similar to Figure \ref{fig:syn}, but convolved with a Gaussian beam with one disc scale height FWHM. The grey contours divide $T_B$ into 4 equally spaced intervals.  }
\label{fig:synsmooth}
\end{figure*}

\begin{figure*}
\includegraphics[width=6.8 in]{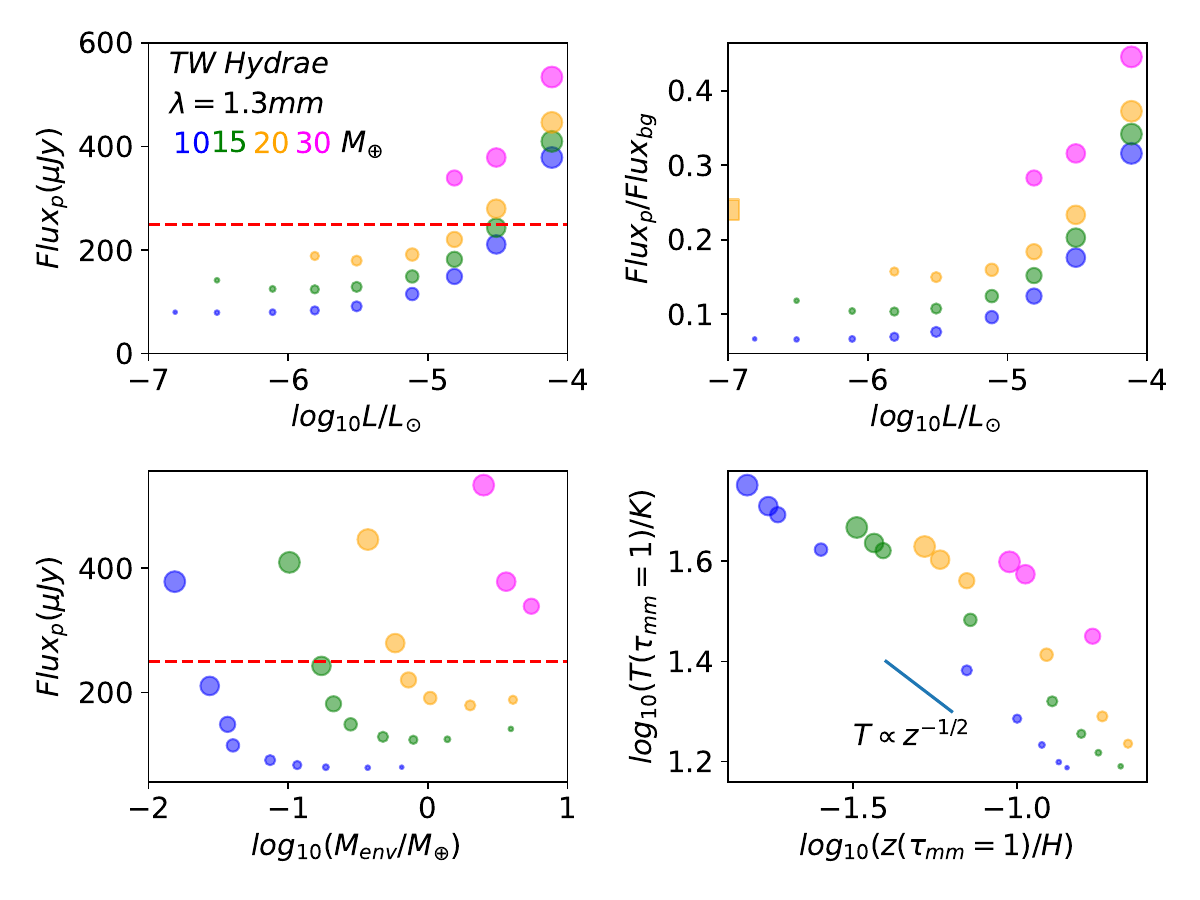}\\
\caption{The upper panels: the flux and flux ratio from the envelope of the potential planet in the TW Hydrae disc with respect to the planet's luminosity.
The bottom panels: the flux with respect to the envelope mass within $r_p$, and the temperature and height at the $\tau_{mm}=1$ surface within the planet's envelope. 
The horizontal line labels the observed flux from the clump (250 $\mu$Jy, \citealt{Tsukagoshi2019}). The bigger points represent cases having higher luminosities. The square
in the upper right panel labels the flux ratio from the 3-D simulation. }
\label{fig:twhydrae}
\end{figure*}

{  We expect good agreements between 3-D and 1-D models, since our 1-D models are tuned to match the density and temperature profiles from 3-D simulations \citep{Zhu2021}. The density and temperature contours in the 3-D simulation (Case C) are shown in Figure \ref{fig:3Dsim}. The Hill radius of the planet is only 0.414 $H$, smaller than $H$. Thus, we don't see any signature of the envelope at the slice of $z=H$, as illustrated in Figure \ref{fig:schematic}. In order to probe the envelope, the whole disc needs to be optically thin. On the other hand, 3-D models reveal that the spirals are present in both the density and temperature contours at all the disc heights, which is similar to previous 3-D global planet-disc interaction simulations \citep{Zhu2015c}. The spiral could be an additional signature of the embedded planet (Section \ref{sec:twhydrae}).  }

The synthetic images from our direct 3-D simulations (case A, B, C) are shown in Figure \ref{fig:syn} and \ref{fig:synsmooth}. Figure \ref{fig:synsmooth} shows the images that are convolved with a Gaussian beam having one disc scale height FWHM. The optical depth of the background disc decreases from the top to bottom panels. At 1.3 mm, the effective optical depth  of the background disc is $\tau_{eff}=$6.4, 2.9, 0.62, while, at 3 mm, the effective optical depth is $\tau_{eff}=$1.7, 1.4, 0.17 from the top to bottom. 

We also integrate the emergent intensity over the area of $\pi$H$^2$ around the planet, and plot the excess flux above the background disc radiation in Figure \ref{fig:fluxrat}. Since the core mass is $\sim$ 20 $M_{\oplus}$ for all these cases, we use big yellow symbols to represent them in Figure \ref{fig:fluxrat}.  
For the case A, B, and C, the planet luminosity  is $\sim10^{-7}$, $10^{-6}$, and $10^{-7}\,L_{\odot}$ at one disc scale height away from the planet center. These luminosities correspond to the x-axis accordingly. Compared with 20 $M_{\oplus}$ cores in 1-D models (small yellow points), these luminosities in 3-D simulations are actually lower than the minimum luminosity the 1-D model allows. As shown in Figure \ref{fig:evolution}, the lowest luminosity for the 20$M_{\oplus}$ core in the 1-D model is $\sim 10^{-6} \,L_{\odot}$. The existence of the minimum luminosity is one important aspect of the core accretion model: decreasing the luminosity increases the envelope mass until reaching the cross-over mass. 
After reaching the cross-over mass, the envelope's self-gravity will lead to run-away accretion which dramatically increases the luminosity. Our 3-D simulations do not consider the envelope's self-gravity, and thus any luminosity is allowed in our 3-D simulations. Furthermore, in our 3-D simulations, the smoothing length of the core's potential is far larger than the real core size, so that the envelope is less massive and its KH contraction releases much less energy. With these caveats in mind, we can see that our 3-D models also show that the planet in the optically thin disc (e.g. case C, or $\lambda=$10 mm for case A/B) has a flux ratio $\gtrsim$ 0.2. When the disc is optically thick (e.g. $\lambda=$1.3 and 3 mm for case A/B),  the flux ratio is significantly lower. For these optically thick cases, the flux ratio from 3-D simulations is higher than those in 1-D models. This could be due to the emission from the spirals, which cannot be captured in 1-D models, as shown below.

The upper left panel in Figure \ref{fig:syn} shows that, when the planet is deeply embedded in the disc and the disc is optically thick, we see little emission from the planetary envelope itself. However, the spirals are much brighter than the envelope. This may not be very surprising since spiral wakes extend vertically throughout the disc and the density perturbation gets stronger towards the disc surface \citep{Zhu2015c}. Even after the convolution, the peak brightness temperature in the spiral is still $\sim$3\% above the background disc brightness temperature. 

Figures \ref{fig:syn} and \ref{fig:synsmooth} also show that when the disc is marginally optically thick/thin (Case A and B), the spectral index 
\begin{equation}
\alpha=\frac{{\rm ln}\frac{T_{B, \nu_1}}{T_{B, \nu_2}}}{{\rm ln}\frac{\nu_1}{\nu_2}} +2\,
\end{equation}
decreases towards the planet position, due to the higher optical depth at the center. When the disc is optically thin, the spectral index is almost a constant across the disc.

\begin{figure*}
\includegraphics[width=6.8 in]{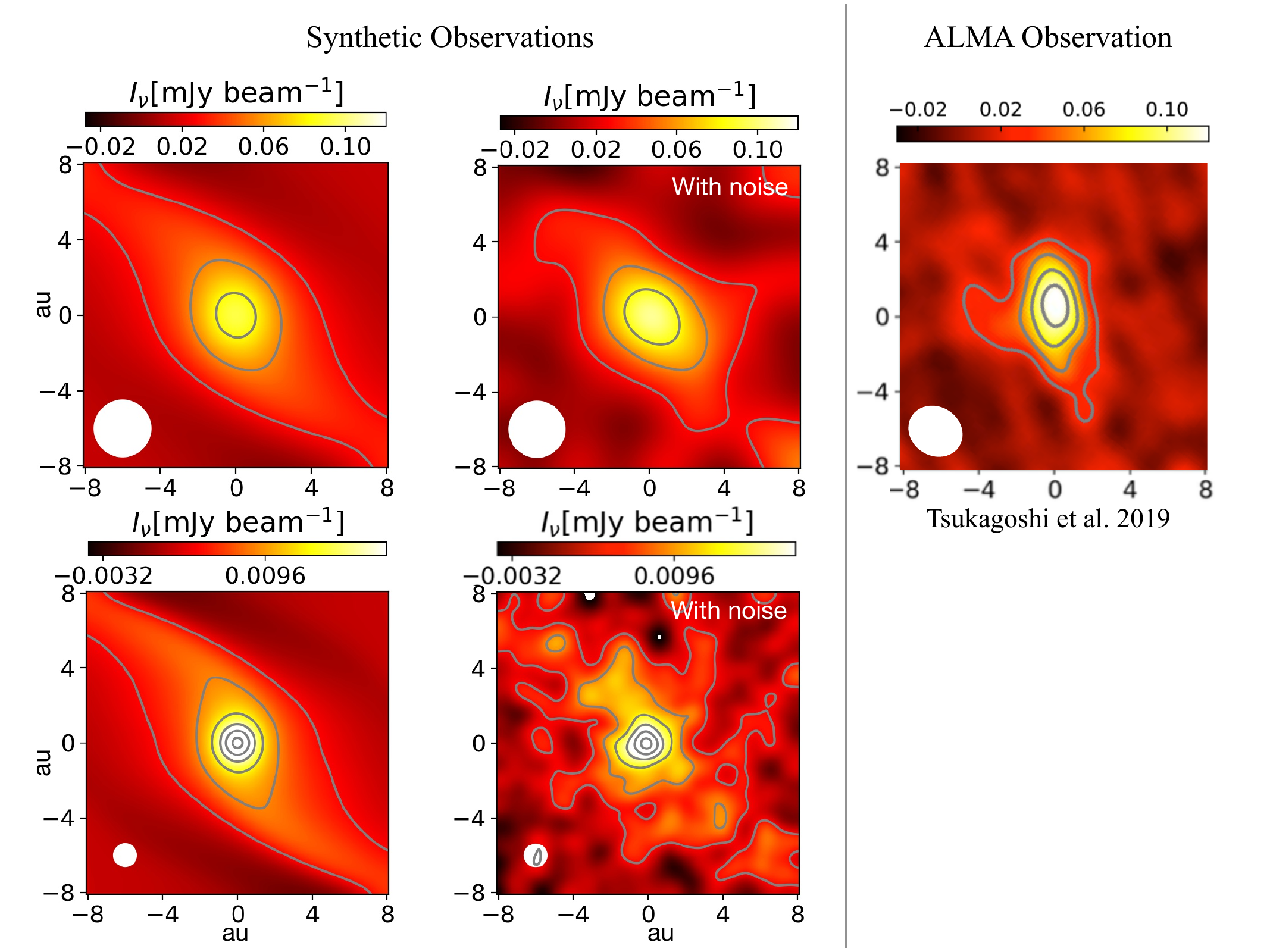}\\
\caption{The left four panels: synthetic images after subtracting the background emission. The images are convolved with a bigger beam (FWHM=0.05'') in the upper panels,
and a smaller beam (FWHM=0.02'') in the bottom panels. The continuum emission is 0.24 mJy/beam in the upper panels and 0.0384 mJy/beam in the lower panels.
Noises are added in the middle panels. The contours in the upper panels are 3, 6,  and 9$\sigma$ with $\sigma$=9.1 $\mu$Jy/beam. The contours in the lower panels are  3, 6, 9, 12, 15, and 18 $\sigma$ with $\sigma$=1.46 $\mu$Jy/beam. The upper right panel is the ALMA observation at band 6 from \citealt{Tsukagoshi2019}. {  The ALMA image has been rotated so that the background flow is in the same direction as that in our shearing box simulation.} }
\label{fig:twhydraesim}
\end{figure*}

\begin{figure*}
\includegraphics[width=6.8 in]{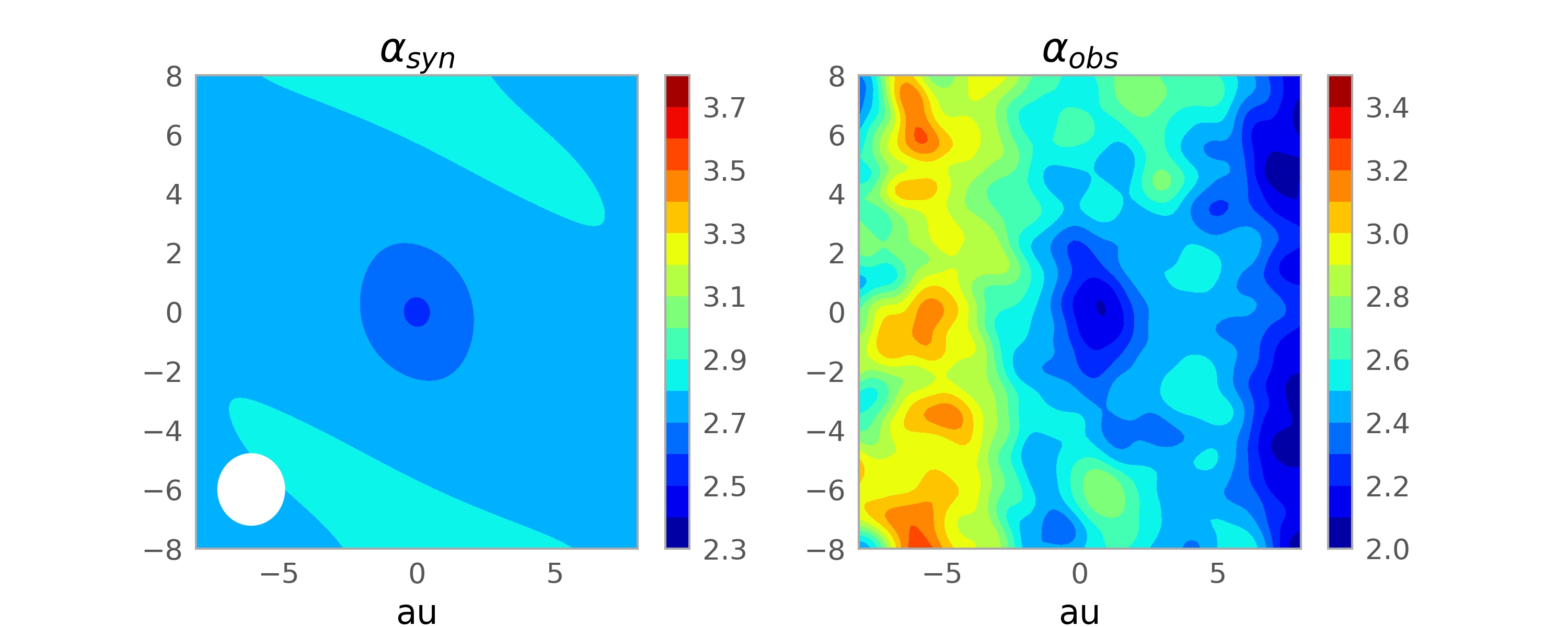}\\
\caption{The left panel: the synthetic image for the spectral index between 0.89 and 1.3 mm from 3-D simulations. The beam size is 40 mas. The right panel: the observed spectral index  between 0.89 and 1.3 mm around the clump \citep{Macias2021}.}
\label{fig:twaalpha}
\end{figure*}

\section{Application to TW Hydrae}
\label{sec:twhydrae}

\cite{Tsukagoshi2019} discovered an au-scale clump at 52 au in the TW Hydrae disc. The integrated excess flux in the 1.3 mm ALMA observation is 250 $\mu$Jy. {  The background disc flux is 122 $\mu$Jy within the area of a Gaussian fit with the major and minor axis of 4.4 and 1 au as the FWHM. Thus, the integrated background flux within the area of $\pi H^2$ is
\begin{equation}
    Flux_{bg}=122 \mu Jy \frac{\pi 3.74^2}{\pi 4.4 \times 1 /(4 {\rm ln(2)})}=1075 \mu Jy\,,
\end{equation}
where we have used Equation \ref{eq:snu} and the disc scale height of 3.74 au at 52 au in TW Hydrae. Thus, the flux ratio is 250/1075=23\%, which is within the range of flux ratios in Figure \ref{fig:fluxrat}. }
This motivates us to examine if a low-mass planet can explain this feature. We adopt the disc structure derived in 
\cite{Macias2021} as our TW Hydrae disc model at 52 au, which is $T_0=15$ K, $\Sigma_0$=11 g cm$^{-2}$ with the dust-to-gas mass ratio of 0.01,
$M_{*}$=0.6 $M_{\odot}$, $\mu$=2.35, and $a_{max}$= 1 mm dust opacity. The resulting 
Rosseland mean opacity at this temperature is 0.387 cm$^2$/g. The disc scale height is 3.74 au. We adopt 59.5 pc for the TW Hydrae's distance \citep{Gaia2016} to calculate the flux. 

We use both 1-D models and 3-D simulations to calculate the radio continuum flux from the planet's envelope.
Figure \ref{fig:twhydrae} shows the planet's excess flux from 1-D models ($Flux_p$ in Section \ref{sec:1demission}).
The observed excess flux (250 $\mu$Jy) from \cite{Tsukagoshi2019} is labeled as the horizontal line. 
Planets with core masses from 10 $M_{\oplus}$ to 20 $M_{\oplus}$ can explain the observations. A higher core mass requires a lower luminosity.
Considering that the planet spends most time at the low luminosity during the KH contraction, the lower luminosity and higher core mass is slightly preferred. 
On the other hand, the KH contraction in the traditional core accretion model is highly simplified (more discussion in Section \ref{sec:discussion}). Together
with the observation uncertainties, the lower end of this mass range ($\sim$10 $M_{\oplus}$) is still possible. The envelope mass in these cases (the lower left panel of Figure \ref{fig:twhydrae}) is much smaller than
the core mass. Thus, the planet's total mass including both the core and envelope is also within this mass range. 

To understand where the thermal radiation is generated inside the envelope, we plot the temperature and the height of the $\tau_{mm}=1$ surface above the planet.  {  As initially discussed in Section \ref{sec:1dmodels},}
when the planet's luminosity decreases (smaller dots in  Figure \ref{fig:twhydrae}), the envelope becomes more massive and the $\tau_{mm}=1$ surface moves higher towards the disc surface (moving to the right in the lower right panel of Figure \ref{fig:twhydrae}). As the luminosity gets lower, the envelope's lower temperature is compensated by the larger emitting area so that the total flux is almost a constant.  
 
Although 1-D models provide some estimates on the envelope's thermal flux, they cannot capture the fluid dynamics occurring around the planet. 
The synthetic images from the direct 3-D simulation are shown in Figure \ref{fig:twhydraesim}. The choices of the dimensionless parameters in Case C  ($\bar{\kappa}=1.7,\beta=10.28$) are based on the TW Hydrae's disc condition. We adopt 0.213 thermal mass for the planetary core mass which corresponds to 20 $M_{\oplus}$. We run the simulations to reach a quasi-steady state. The derived luminosity of the envelope increases from $10^{-8}$ $L_{\odot}$ at the center to $10^{-7}$ $L_{\odot}$ at the scale height away from the center. {  We derive $Flux_{p}/Flux_{bg}$ =0.23 from the 3-D simulations, which is shown in the upper right panel of Figure \ref{fig:twhydrae}. This ratio is the same as 0.23 derived form the ALMA measurement (beginning of Section \ref{sec:twhydrae}). We can also compare the synthetic image with the ALMA observation. }
The left two panels in the top row of Figure \ref{fig:twhydraesim} are plotted similarly to the plot in \cite{Tsukagoshi2019} (shown in the upper rightmost panel), with the same spatial scale, color range and contour levels. The images are convolved with a Gaussian beam having FWHM=50 mas, similar to the beam size in the observation. 
The background emission (0.24 mJy/beam) has been subtracted, and a noise of $\sigma\sim$9.1 $\mu$Jy/beam (similar to the $\sigma$ in observations) is added in the middle panel. We  rotate and then flip the observational image in \cite{Tsukagoshi2019} so that the star is towards the right side of the image and the disc background flow is in the same direction as the flow in the shearing box simulation. The resulting synthetic image has similar brightness and morphology as the ALMA observation. The envelope is marginally resolved due to several reasons. First, the Hill radius is around half of the disc scale height. Thus, the high density envelope is quite extended. Second, the background disc is marginally optically thick, leading to a more uniform intensity in the extended region around the planet (as shown in Figure \ref{fig:syn}). Third, the contribution of the spirals elongates the emitting region. To fully resolve this clump, we need to carry out higher resolution observations, as demonstrated in the bottom panels of Figure \ref{fig:twhydraesim}.

The spirals in the middle panels of Figure \ref{fig:twhydraesim} break up and are barely visible. Despite this, the simulated spirals seem to be a little bit more visible than those in observations. While we are preparing this manuscript, new analysis by \cite{Tsukagoshi2022} suggests that there may indeed be a tail (spiral) connected with the clump in the observation, which agrees better with our simulation.  With a higher spatial resolution and sensitivity (bottom panels of Figure \ref{fig:twhydraesim}), both the clump and the spiral will become more resolved and apparent. 

To test our model, we can compare the predicted intensity at another wavelength with observations, or compare the spectral index between observations at these two wavelengths. Fortunately, TW Hydrae has been observed at multiple wavelengths. We compare the spectral index map  between ALMA Band 6 and 7 observations \citep{Macias2021} with our model prediction in Figure \ref{fig:twaalpha}. As mentioned in Section \ref{sec:3dmodels}, when the envelope is marginally optically thick ($\tau_{eff}=0.5$ for the TW Hydrae clump), the clump center has a lower spectral index due to the increase of the optical depth. Observations clearly show this decrease of $\alpha$ towards the clump center. Observations and theory agree reasonably well. 

We note that, besides the planetary envelope, several other physical processes may also explain the observed au-scale excess. First, it can be a small vortex in the disc. Anticyclonic vortices can trap dust particles \citep{Barge1995}. This excess flux of 250 $\mu Jy$ is equivalent to 0.047 $M_{\oplus}$ dust  using our dust opacity.  On the other hand,  vortices that are smaller than the disc scale height may be subject to the elliptical instability \citep{Lesur2009}. Numerical simulations  show that small vortices normally merge to form large vortices that can be long-lived \citep{Shen2006}. 
These large vortices are proposed to explain the large scale disc asymmetry \citep{vandermarel2013} that is very different from this au-scale clump. Although a vortex can also excite spirals through the interaction with the supersonic region \citep{Paardekooper2010}, the spirals are much weaker \citep{Huang2018}. We expect weaker spirals from a small vortex that is well within the supersonic region. The second proposed mechanism is the dust emission from a dust-losing planet \citep{Nayakshin2020}. This mechanism is motivated by the elongated dust shape in the observation. However, the reason for a planet to lose a large amount of dust is unclear. \cite{Nayakshin2020} suggests that it could be planet-planet collision or the destruction of a planet formed in a gravitationally unstable disc. 

On the contrary, our hydrostatic envelope explanation is much simpler and follows the traditional planet formation model. In fact, all disc and dust parameters are from ALMA constraints in \cite{Macias2021}, and the only parameter we add is the 10-20 $M_{\oplus}$ planetary core in the disc. The synthetic image reproduces the flux, the spectral index dip, and the slightly elongated structure.

\section{Discussion}
\label{sec:discussion}
\subsection{A Simple Model} 
\label{sec:simplemodel}
\begin{figure}
\includegraphics[width=3.3 in]{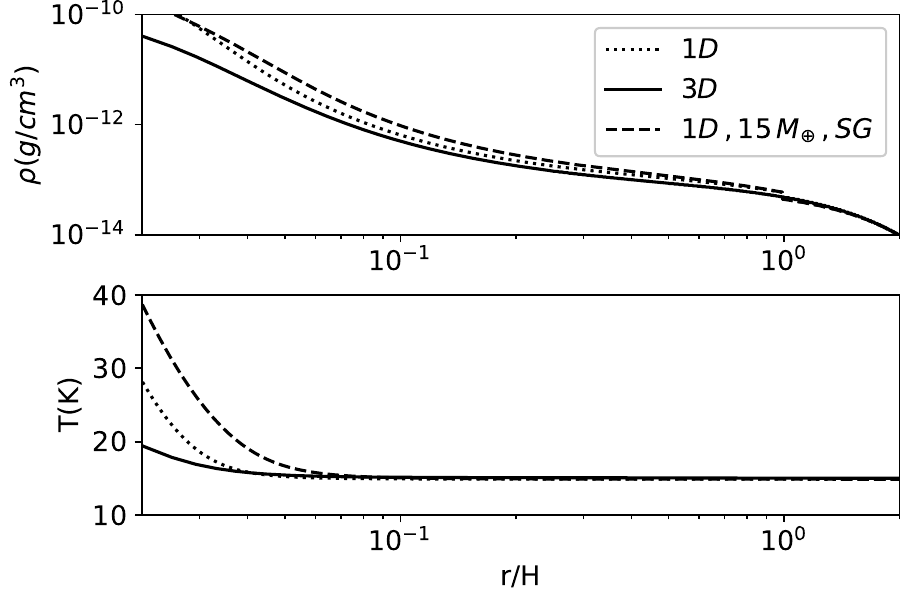}\\
\caption{The radial profiles of the planet envelopes for different models to simulate the clump in the TW Hydrae disc. The models labeled with ``1-D'' and ``3-D'' have a 20 $M_{\oplus}$ core and do not consider
the envelope's self-gravity.  The other 1-D model has a 15 $M_{\oplus}$ core and has considered the envelope's self-gravity.}
\label{fig:rprofile}
\end{figure}

To get insights into the envelope's thermal emission, we plot the radial profiles of the planet envelope in different models for the TW Hydrae clump, shown in Figure \ref{fig:rprofile}. 
The 3-D simulation has a smoothing length of 0.0212 H (0.1 $R_{B}$), which explains the density and temperature deviation at that scale. 
We  carry out two different 1-D models, with and without considering the envelope's self-gravity. 
For the model with the envelope's self-gravity,
the lowest allowed luminosity is higher than the luminosity in models without considering self-gravity (discussed in Section \ref{sec:3dmodels}). Thus, the self-gravity model has a faster temperature rise towards the center.
To get a similar density profile as other models with a 20 $M_{\oplus}$ core, 
we also have to adopt a smaller core mass (15 $M_{\oplus}$) in this self-gravity model to compensate  for the effect of the envelope's gravity. 

For all these models, the envelope's density increases towards the center and it is isothermal when $r/H\gtrsim$ 0.05. With the isothermal assumption,
we can integrate Equation \ref{eq:Pr} to derive
\begin{equation}
\frac{\rho(r)}{\rho(r_z)}=e^{\frac{GM_p}{c_{s}^2}\left(\frac{1}{r}-\frac{1}{r_z}\right)-\frac{(r^2-r_z^2)}{2 H^2}}\,,\label{eq:rhorz}
\end{equation}
where $GM_p/c_s^2$ is the Bondi radius. Thus, we have
\begin{equation}
\rho(r)=\rho(H)e^{\left(\frac{r_B}{r}-\frac{r_B}{H}\right)-\frac{(r^2-H^2)}{2 H^2}}\label{eq:rhor}\,,
\end{equation}
and we can also assume that $\rho(H)\approx \rho_{mid}exp(-1/2)$, where $\rho_{mid}$ is the disc's midplane density
without the planet.
Within the Hill radius, the first term in the exponent dominates. Furthermore, when $r\ll r_{B}$ and $r\ll H$, we have $\rho(r)=\rho_{mid}exp\{r_{B}/r\}$. If we integrate the optical depth radially, $\int\rho(r)\kappa_{mm}dr$,
we can derive the radius ($r_{mm}$) where the envelope becomes optically thick at the observed wavelength. To be more accurate, we use the effective optical depth considering the scattering.
For the TW Hydrae case, we have $r_{mm}\sim$0.035 $H$. 
Then, we can integrate the density to derive the mass difference between the envelope and the background disc within the same region
\begin{equation}
\Delta M=\int_{0.035 H}^H (\rho(r)-\rho_{mid}e^{-r^2/2H^2})4\pi r^2 dr\,,\label{eq:deltaM}
\end{equation}
with $\rho(r)$ from Equation \ref{eq:rhor} and $r_B=0.212 H$. Numerically, we derive $\Delta M=0.55 \rho_{mid}H^3$.
On the other hand, the background disc mass within H is $\pi H^2 \sqrt{2\pi} H \rho_{mid}\sim 7.87 \rho_{mid}H^3$.
 Considering that this region is isothermal and the disc is optically thin (the effective optical depth of the background disc is 0.5),  
the ratio between these two masses ($\sim7\%$) equals the flux ratio between the planet and the background disc. This ratio is similar to
the 10-20\% measured in 1-D and 3-D models as shown 
in the upper right panel of Figure \ref{fig:twhydrae}. If we consider that the effect of the planet's gravity extends beyond 
$H$ (e.g. Equation \ref{eq:Hmid}) or simply use 2$H$ as the integration limit in Equation \ref{eq:deltaM}, we can double the mass ratio estimate, which is even closer to the ratio in simulations. On the other hand, the envelope is not symmetric beyond $H$ and 
the structure there can only be captured in 3-D simulations.

Our above calculation provides a simple way to estimate the envelope's excess flux for any given disc and planet properties, as long as the envelope is almost isothermal and the disc is optically thin. It also explains why the flux ratio will flatten out to non-zero values when the luminosity is low (as shown in the upper right
panel of Figure \ref{fig:twhydrae} and Figure \ref{fig:fluxrat}). Even if the planetary envelope is isothermal with an extremely low luminosity, the envelope density in the region
within $H$ still increases according to Equation \ref{eq:rhorz}. This extra envelope mass beyond the $\tau_{mm}=1$ surface 
can be a moderate fraction of the local disc mass. When the disc is optically thin, we will be able to probe this extra mass. 

\subsection{Caveats}
\label{sec:caveat}
\begin{figure*}
\includegraphics[width=6.8 in]{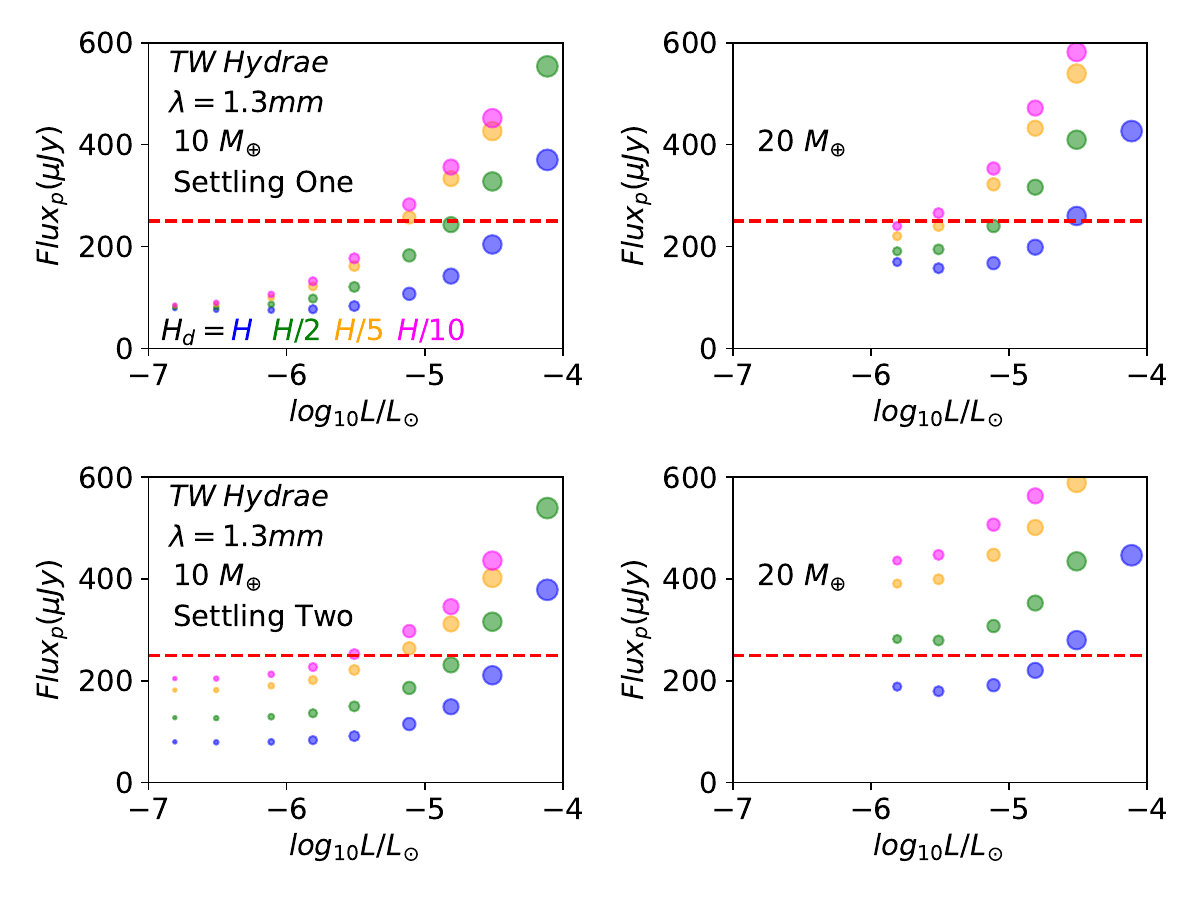}\\
\caption{The 1.3 mm flux of the potential planet in the TW Hydrae disc with respect to the planet's luminosity considering that dust is settled in the disc. The left panels are with 10 $M_{\oplus}$ cores, while the right panels are with 20 $M_{\oplus}$ cores.
Two different treatments for dust settling (Equations \ref{eq:rhod1} and \ref{eq:rhod2}, upper and lower panels) have been considered. Different colors represent different thickness of the dusty disc. }
\label{fig:twhydraesettle}
\end{figure*}

Since the inner envelope is optically thick, the envelope's radio thermal emission is insensitive to the
processes happening in the deeper part of the envelope. No matter if the planet is built through pebble accretion \citep{Brouwers2018,Brouwers2020}
or planetesimal accretion, the outer envelope is quite similar and only determined by the luminosity, the opacity and the core mass.
On the other hand, several processes in the disc can affect the envelope's thermal emission. 

The first process is dust settling. {  Dust settling can affect both the envelope's thermal structure and its radio emission. Dust settling can reduce the Rosseland mean opacity at the disc surface and the planet's outer envelope, while increasing the opacity at the disc midplane and the inner envelope. This results in decreasing the temperature gradient ($|dT/dr|$) at the outer envelope and increasing the gradient at the inner envelope. Simulations considering dust settling need to be constructed to study its effect on the envelope's thermal structure \citep{Krapp2021}. Without carrying out such simulations in this work, we will assume that the envelope's thermal structure is unaffected by the dust settling (e.g. in the scenario where small unsettled dust dominates the Rosseland mean opacity) and just focus on studying how settled mm-sized particles affect the envelope's radio emission.}
If mm-sized dust is significantly settled, the $\tau_{mm}=1$ surface will be more likely within the planet's Hill sphere, so that the high density and temperature there can boost the planet's thermal emission. To quantify the effect of dust settling, we  modify our $\bar{r}-z$ grid that is generated from the 1-D structure calculation to allow for dust settling. We use the same gas and temperature structure as before, but modify the dust distribution in the grid in two different ways.  In the first approach, we assume that dust is settled vertically in the disc. We integrate the density in the $\bar{r}-z$ grid vertically along $z$ to derive the surface density profile ($\Sigma(r)$). Then, we reassign the grid density as
\begin{equation}
\rho_d(\bar{r},z)=\frac{\Sigma(\bar{r})\epsilon}{\sqrt{2\pi}H_d}e^{-\frac{z^2}{2H_d^2}}\,,
\label{eq:rhod1}
\end{equation}
where $\epsilon=0.01$ is the dust-to-gas mass ratio, $H_d$ is the dust scale height and we vary $H_d$ from $H/10$ to $H/2$. We can think of this approach to represent that the dust and gas gather around the planet and then dust settles to the disc midplane. Fluxes from the planetary envelopes are given in the upper panels of Figure \ref{fig:twhydraesettle}. In this approach, when the luminosity is low, settling has little effect on the dust emission.  The temperature in the envelope equals the background disc temperature, and the vertically integrated density is also unchanged. When the luminosity is high, stronger settling increases the emission. This is because {  the temperature at the inner envelope} is significantly higher than {  the temperature at the outer envelope} in these luminous cases, and with stronger settling {  the $\tau_{mm}=1$ surface reveals the hotter inner envelope}.  In the second approach, we assume that the dust density increases proportionally to the gas density in the envelope. The proportionality is the background dust density change due to settling without the presence of the planet. Thus, we have
\begin{equation}
    \rho_d(\bar{r},z)=\rho(\bar{r},z)\epsilon\frac{H}{H_d}e^{-\frac{z^2}{2}(\frac{1}{H_d^2}-\frac{1}{H^2})}\,.\label{eq:rhod2}
\end{equation}
We can also think of this approach to represent gathering-and-settling, but in an opposite way from Equation \ref{eq:rhod1}. In this approach, dust first settles in the disc and then the planet gathers the envelope (including both gas and dust) horizontally. Equation \ref{eq:rhod2} is also consistent with the density structure expected from pebble accretion as described in Section \ref{sec:formation}. {  Equation \ref{eq:rhod2} leads to a higher dust density than Equation \ref{eq:rhod1} within the envelope since $\rho(\bar{r},z)$ in the envelope is always larger than the background disc density $\Sigma(\bar{r})/(\sqrt{2\pi}H)exp\{-z^2/(2H^2)\}$. }
Thus, this settling model provides higher fluxes (the resulting fluxes in the bottom panels of Figure \ref{fig:twhydraesettle}), and the envelopes around a 10 $M_{\oplus}$ core can explain the TW Hydrae observations. To understand which model is more realistic, we need to include dust in 3-D simulations as in \cite{Krapp2021}. Overall, our simple models suggest that settling increases the envelope's thermal flux. 

The second process which can affect the radio emission is the additional heating generated by the spirals. 
As shown in 3-D simulations, the spirals can be even brighter than the planet itself when the planet is hidden inside the disc.
The high temperature
within the spirals  increases
the thermal emission from the region around the planet. 

The third process that can complicate the picture is the gap opening by the planet. Within the gap,
the disc density can decrease significantly. {  Although the envelope's growth is quite insensitive to the ambient disc density (planetary accretion depends on the disc density logarithmically, as in \citealt{LeeChiang2016,Ginzburg2016}), the envelope's $\tau_{mm}=1$ surface (and thus its radio emission) is sensitive to the background disc density.} Both the disc and envelope
emission can be weak within the gap, unless a massive circumplanetary disc forms around the planet \citep{Zhu2018}. 
However, for a low-mass planet, the gap opening timescale can be very long.
If we divide the angular momentum of an annulus having the width of $H$ (which is $2\pi R H\Sigma \Omega_p R^2$) with the angular momentum flux excited
by the planet (which is $\sim (GM_p)^2\Sigma R_p\Omega_p/c_s^3$, \citealt{Goldreich1980}), we can derive the gap opening timescale as
\begin{equation}
t_{gap}=\left(\frac{M_{th}}{M_p}\right)^2\left(\frac{R}{H}\right)^2 t_{orb}\,.
\end{equation}
A 10 $M_{\oplus}$ planet at 52 au in the TW Hydrae disc is equivalent to a  0.1 $M_{th}$ planet in a $H/R$=0.078 disc, which leads to the gap opening
timescale of 1.6$\times10^4$ local orbits or 7.8 Myrs, longer than a typical disc's lifetime. Thus, the potential planet is unlikely to induce a deep gap in the TW Hydrae disc, consistent with the observations. We notice that there are very shallow gaps/plateaus at 49 and 58 au in the submm intensity profiles of TW Hydrae (Figure 7 of \citealt{Macias2021}), which could be shallow gaps induced by the planet. 

\subsection{Implications to Planet Formation}
\label{sec:formation}
If the clump in the TW Hydrae disc indeed reveals an envelope around a 20 $M_{\oplus}$ core (although a 10 $M_{\oplus}$ core is still possible if we consider dust settling as shown above), it is in tension with the 10 $M_{\oplus}$ critical core mass based on the traditional core accretion model (Figure \ref{fig:evolution}). Several factors can alleviate this tension. First, since  it takes $\sim$ 0.2 Myrs for a planet with a 20 $M_{\oplus}$ core to reach the run-away phase (Figure \ref{fig:evolution}), this core in TW Hydrae could be formed recently within the past 0.2 Myrs. This timescale is not very short compared with a protoplanetary disc's lifetime.  This scenario implies that planet formation is a continuous process during a disc's lifetime. {  Second, a higher disc opacity could also slow down the envelope accretion \citep{Pollack1996,Ikoma2000,Chachan2021}. { Such opacity increase could be due to dust fragmentation at ice lines, dust concentration within gaseous rings, or some other effects.}}
Third, additional heating sources may affect the envelope's KH contraction. As shown in the luminosity curve in Figure \ref{fig:evolution}, an additional luminosity of $10^{-6}$-$10^{-5}$ $L_{\oplus}$ (e.g. from solid accretion) will be larger than the luminosity due to the KH contraction. For a 20 $M_{\oplus}$ core with $r_{c}=3 R_{\oplus}$, this luminosity ($L=GM_c\dot{M}/r_{c}$) corresponds to $\dot{M}$ of 5$\times 10^{-6}$-5$\times 10^{-5}$ $M_{\oplus}/yr.$ This accretion will increase the planet mass by 5-50 $M_{\oplus}$ over 1 Myrs, which is still moderate compared with the 20 $M_{\oplus}$ core. {  On the other hand, it is unclear if such additional accretion would shorten or delay the run-away accretion, since accretion not only releases the energy but also increases the core mass \citep{LeeChiang2015}. }

One way for this planet to accrete is through pebble accretion \citep{Ormel2010,Lambrechts2012}, especially considering that this planet is indeed embedded in a dusty disc with mm particles. Considering that mm dust is highly settled and the shear velocity dominates the headwind velocity for this massive core, we use the 2-D pebble accretion rate in the shear regime:
\begin{equation}
    \dot{M}\sim 2R_{H}^2\Omega T_s^{2/3} \Sigma_d \,,\label{eq:ts}
\end{equation}
where $T_s=t_s/\Omega$ is the dust's dimensionless stopping time \citep{Ormel2017}, and $t_s$ is the stopping time
\begin{equation}
    t_s=\frac{s\rho_s}{\sqrt{8/\pi}\rho_g c_s}
\end{equation}
in the Epstein regime \citep{Weidenschilling1977} with $\rho_s$ as the solid density of the particles.
With the gas surface density of $\Sigma_0$=11 g cm$^{-2}$, a 1 mm particle has $T_s=1.4\times 10^{-2}$. If $\Sigma_d=\Sigma_0/100$, we can calculate the pebble accretion rate as 1.7$\times 10^{-5}$ $M_{\oplus}$/yr. Over 1 Myrs, it will accrete 17 $M_{\oplus}$, similar to its current mass.
This accretion will also generate $L\sim 3.5\times10^{-6}$ $L_{\odot}$, which is roughly consistent with the needed luminosity in Figures \ref{fig:twhydrae} and \ref{fig:twhydraesettle}. Thus, both the mass accretion rate from pebble accretion and the generated luminosity are not in conflict with our proposed 10-20 $M_{\oplus}$ planet in the disc to explain the ALMA observations. 

If the planet is accreting through pebble accretion, it may affect the dust distribution within the envelope. Although a proper study on this needs 3-D simulations with both gas and dust components, we can do some rough estimates on the effect. The impact parameter for the planet in the shear regime is $b\sim T_s^{1/3}R_H$ which is 0.11 $H$ in this case. Only dust distribution within the impact parameter will be affected by pebble accretion. Considering the $\tau_{mm}=1$ surface is at $r$=0.035$H$ as derived in Section \ref{sec:simplemodel}, radio observations will only be able to probe the very outskirt of the pebble accretion region. Density there will not be significantly affected by pebble accretion. Even so, we can still estimate the density within the impact parameter. Due to mass conservation in a steady state, we have
$\rho_d(r)4\pi r^2 v_r=\dot{M}$ or $\Sigma_d(r)2\pi r v_{r}=\dot{M}$ depending on if the dust at $r<b$ is spherically distributed or still concentrated at the midplane.  $v_r$ is the dust's settling velocity.  Since we focus on the small region inside $b$, we assume that dust distribution is more spherically distributed. The settling velocity $v_r$ equals $g t_s$ where $g=GM_{p}/r^2$.  If we plug $t_s$ (Equation \ref{eq:ts}) and $v_r$ into the mass conservation equation, we have
\begin{equation}
    \rho_d(r)   = \frac{\sqrt{8/\pi}\dot{M}\rho_g c_s}{4\pi G M_p s \rho_s}\,,
\end{equation}
where $\dot{M}$, $M_p$, $s$, and $\rho_s$ are all constant. In an isothermal envelope where $c_s$ is also a constant, we have $\rho_d(r)\propto \rho_g(r)$, which means that the dust-to-gas density ratio is a constant within the impact parameter. This is also consistent with Equation \ref{eq:rhod2} in Section \ref{sec:caveat}. Nevertheless, pebble accretion cannot change the envelope's dust distribution significantly, at least at $r>b$.

{  Finally, if this clump in TW Hydrae is indeed the envelope of a low mass planet, it suggests that low-mass planets could be common in protoplanetary disc. As discussed in the next section, TW Hydrae is the only system that we can confidently detect the envelopes around low-mass planets with a moderate ALMA integration time. Finding a low-mass planet candidate in the only system where we can potentially detect such planets implies that these low-mass embedded planets could be common in protoplanetary discs. { On the other hand, one example is not statistically significant and more deeper ALMA observations are needed in future.} }

\subsection{Using ALMA and ngVLA to Constrain Planet Formation}
\label{sec:ngvla}
It is quite challenging to detect the low-mass planet even with ALMA observations.
{  TW Hydrae is one of the closest protostar ($\sim$59.5 pc), making such detection possible. } The observation by \cite{Tsukagoshi2019} has a spatial resolution of $\sim$ 2.5 au, smaller than the 4 au disc scale height at 52 au. Together with the high sensitivity ($\sigma\sim$9 $\mu$Jy/beam) at 1.3 mm, the 250 $\mu$Jy clump can be robustly detected. Since other protoplanetary discs are much further away ($\gtrsim$100 pc), it is difficult to obtain similar resolution and sensitivity. The closest observations regarding resolution and sensitivity are DSHARP observations \citep{Andrews2018}, which have resolutions of $\sim$5 au at 150 pc and $\sigma\sim$10-20 $\mu$Jy/beam at 1.3 mm. Several potential point sources with $\sim$50-100 $\mu$Jy have been found in these discs \citep{Andrews2021}. \cite{Andrews2021} search
for circumplanetary discs (CPDs) within the gaps. Unlike CPDs whose thermal radiation largely depends on the CPD accretion physics \citep{isella2014,Zhu2018}, the flux from the embedded planetary envelope studied here should roughly scale with the background disc flux when the disc is optically thin. The disc serves as the outer boundary of the envelope, so that the density at the outer envelope scales with the disc density (Equations \ref{eq:rhorz} and \ref{eq:rhor}).
The background disc flux per beam at 1.3 mm is
\begin{equation}
    Flux_{bg}=335 \frac{T_B}{10 K}\left(\frac{\theta}{33\, mas}\right)^2 \mu Jy\,,\label{eq:fluxbgdsharp}
\end{equation}
where $T_B$ is the disc's brightness temperature, and $\theta$ is the beam width. The 5 au resolution in \cite{Andrews2018} is  probing the disc region at the scale of $H$ for the outer disc beyond 50 au. Thus,
we can compare the observed flux ratio (the clump flux divided by the background disc flux (Equation \ref{eq:fluxbgdsharp})) with those calculated in Figure \ref{fig:fluxrat} to study if these clumps can be low-mass planets. 
Here, we focus on the ring/disc region instead of the gap region since low-mass planets should not induce deep gaps. Most disc regions in DSHARP discs beyond 40 au have $T_B$ from 1 to 10 $K$ \citep{Huang2018,Andrews2021}. Thus, the potential clumps of 50-100 $\mu$Jy  have flux ratios $\gtrsim$15\%, comparable with those from 20 $M_{\oplus}$ cores (Figure \ref{fig:fluxrat}). In future, it is worth carrying out higher resolution and more sensitive observations at multiple wavelengths to confirm these observed clumps. 

Based on the physical pitcutre presented, we can develop an observational strategy to search for the envelopes around low-mass planets. First, we need to observe the discs that are slightly optically thin. If the disc is too optically thin, the flux from both the disc and the envelope is low and hard to detect, since the envelope's flux roughly scales with the disc flux. If the disc is too optically thick, the high density region within the Hill sphere is hidden from the observation and the flux ratio between the envelope and the background disc is small (Figure \ref{fig:fluxrat}). Second, both high spatial resolution and  sensitivity are desired. To detect the envelope around a 10 $M_{\oplus}$ core, we need to confidently detect a clump that is 10\% brighter than the background disc with a resolution smaller than the disc scale height. Practically, although a very high resolution observation  allows us to study the shape of the emitting region, it also requires significantly more observing time. Thus, marginally resolving the clump with several beams could be ideal. Finally, to distinguish a true detection from a random noise, we also need additional evidences. The first additional evidence could be from observations at different wavelengths. Multi-wavelength observations can also test if the spectral index at the clump is consistent with the theoretical expectation. The second additional evidence could be the detection of spirals. The spirals excited by Neptune mass planets may be detectable with ALMA \citep{Speedie2022}. From these perspectives, the TW Hydrae clump is ideal for detection and may indeed be the envelope of an embedded planet because: 1) $\tau_{eff}\sim$ 0.5, 2) the observation resolves half the disc scale height, 3) the spectral index is consistent with the theory prediction, and 4) the spirals around the planet may have been detected already.

\begin{figure*}
\includegraphics[width=6.8 in]{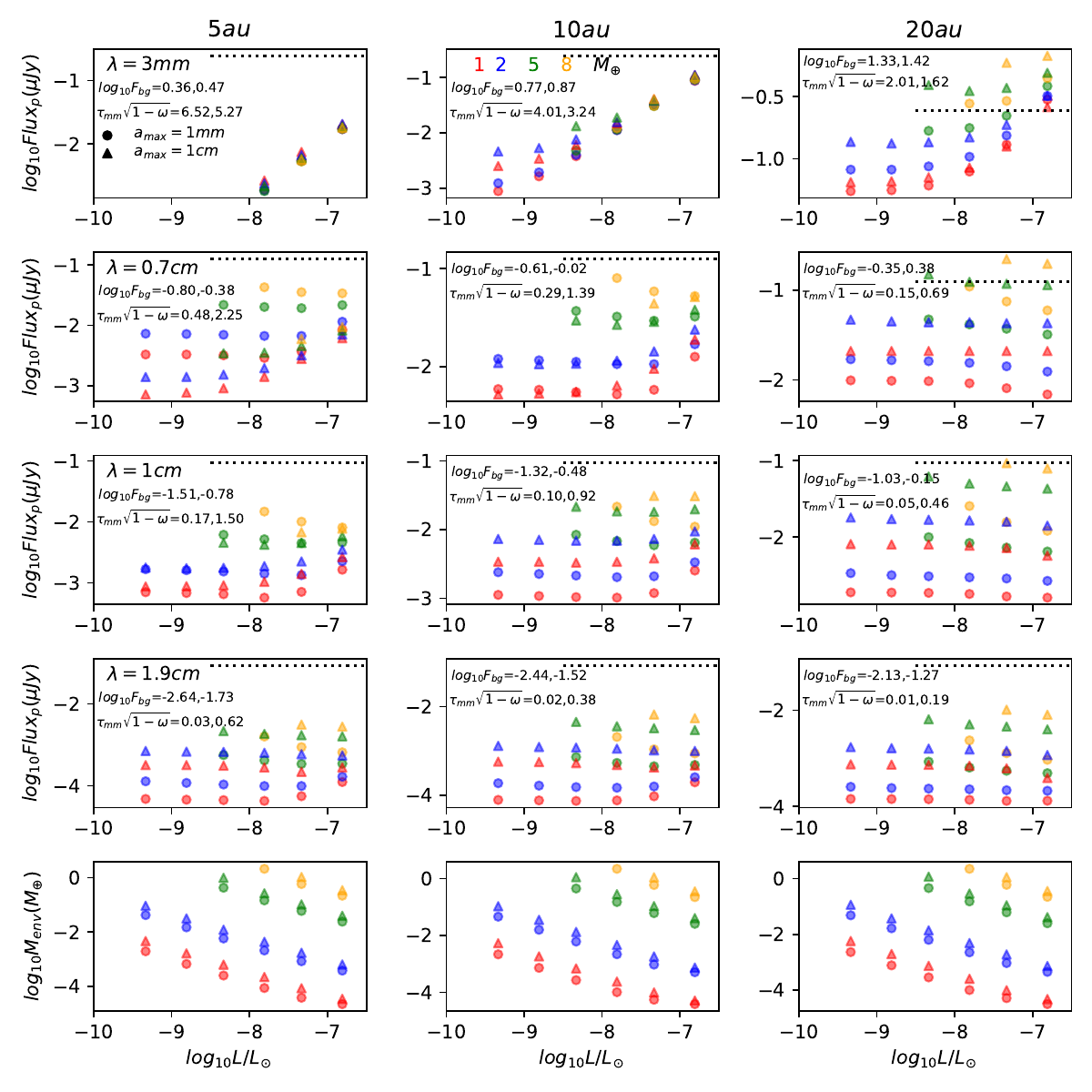}\\
\caption{Similar to Figure \ref{fig:1mm} but for planets at smaller disc radii and being observed at longer wavelengths (ngVLA bands). The dots represent the cases using the opacity with $a_{max}$=1 mm dust in the disc, while the triangles use the opacity with $a_{max}$=1 cm dust in the disc. There are two values for $F_{bg}$ and $\tau_{eff}$ in each panel. The first one corresponds to $a_{max}$=1 mm dust, while the second one corresponds to $a_{max}$=1 cm dust.  {The horizontal lines are 5$\sigma$ detection limits of ngVLA with 100 hr on-source integration and 10 mas resolution at each band.}}
\label{fig:fluxratngvla}
\end{figure*}

In future, we would like to find lower mass planets at  inner discs. Since the inner discs are more optically thick, we need to observe at longer wavelengths. To find lower mass planets, we also need observations with a higher sensitivity. Fortunately, ngVLA will meet both requirements. Figure \ref{fig:fluxratngvla} shows the envelope's radio flux at the ngVLA bands for low-mass planets at the inner discs.  To resolve the disc scale height ($\lesssim$1 au) at the inner disc, the resolution needs to be $\lesssim$ 10 mas for a source 100 pc away. Based on the ngVLA specs \footnote{\url{https://ngvla.nrao.edu/page/performance}}, 0.3 cm, 0.7 cm and 1 cm observations with the 10 mas resolution have the rms of 0.49, 0.25 and 0.19 $\mu$Jy/beam per hour integration. The top three rows in Figure \ref{fig:fluxratngvla} show that, with 100 hours integration, we may be able to find $\gtrsim$ 8 $M_{\oplus}$ planets at $\gtrsim$ 10 au. 


The low-mass planets (1-20 $M_{\oplus}$) should be abundant based on both core-accretion theory and exoplanet statistics. Although it is challenging to detect low-mass planets, we should not be surprised to find them in protoplanetary discs. If higher-mass planets ($\gtrsim$ Neptune mass) are responsible for most gaps/rings in protoplanetary discs, the ubiquitous rings in discs suggest that the higher-mass planets are also abundant. If we can constrain the low-mass planet population and compare it with the high-mass population, we can potentially reconstruct the planet evolutionary history. In the protostellar studies, a similar approach was used to constrain that the protostellar infall phase lasts 10$^5$ years since 10\% of the total pre-main-sequence population ($\lesssim$1 Myr old) is protostars \citep{Beichman1986,Kenyon1990}. 

More specifically for the ``protoplanet'' population, we can constrain the fraction of the disc lifetime ($f_{KH}$) a planet will spend before reaching the run-away accretion phase. For example, if the embedded low-mass planets are 9 times more abundant than massive planets in protoplanetary discs, we can derive $f_{KH}\leq$90\% assuming that planets are born continuously during the disc's lifetime.
The less sign is due to that not all low-mass planets will become giant planets even if they are born at the beginning of the disc formation (e.g. with a low mass core).
If only 1/3 of the embedded planets are massive enough to be capable of becoming giant planets, we have $f_{KH}$=75\%. On the other hand, if embedded low-mass planets are not discovered in protoplanetary discs, high-mass planets in discs may not grow from low-mass planets, and they may form directly through gravitational instability.

\section{Conclusion}
\label{sec:conclusion}
In this paper, we use both 1-D models and 3-D simulations to study the KH contraction of planetary envelopes and the observational signatures of these envelopes. 
Based on the traditional core-accretion theory, the KH contraction phase is the bottleneck of giant planet formation.
The planet spends most of its time during this phase, trying to reach the cross-over mass for run-away accretion.
With our adopted disc structure and opacity, our 1-D evolutionary model suggests that
the planet evolutionary track is almost identical from 20 au to 100 au, and the critical core
mass for reaching the run-away phase is a constant ($\sim$10 $M_{\odot}$) at different disc radii. 

Considering that this KH contraction is the longest phase of a planet's evolution, there could be a large population of low-mass
embedded young planets at this phase unless the initial core mass is much larger than 10 $M_{\odot}$. To study if we can detect such planets, we use 1-D models and 3-D simulations 
to calculate the envelope structure 
around the planetary cores (with several to tens of $M_{\oplus}$) having different luminosities, and derive their
thermal fluxes at radio wavelengths. When the background disc is optically thick at the  wavelength of the observation 
($\tau_{eff}\gtrsim$1), the flux of the embedded planet only depends on its luminosity 
(almost has a linearly relationship) and has no
dependence on the planet mass. When the background disc is optically thin, we can see through the disc and
probe the denser envelope within the planet's Hill radius. At a given luminosity, the envelope around a more
massive core produces stronger thermal emission. Although the radio thermal emission decreases with a smaller planet luminosity similar
to the optically thick case,  the thermal emission flattens out or even increases when the luminosity decreases to very low values. 
This is because a low luminosity leads to a cooler envelope which is also denser and more massive. 
Even in an envelope that is isothermal with the disc temperature, the additional mass gathered within the Hill sphere due to the planet's gravity is still significant.

Since the planets are embedded in the disc, it is the flux ratio between the planet's thermal emission and the background disc radiation
that determines if we can detect these planet.  We find that sufficiently deep and high spatial resolution radio observations, 
e.g. reaching 10\% sensitivity above the background flux over the area of $\pi H^2$, can reveal the envelopes around 10$M_{\oplus}$ planets
when the background disc is optically thin.  Dust settling increases the flux ratio even further, making the envelope more apparent.
The 3-D simulations suggest that the planet-induced spirals may also be detectable, and the spectral index decreases
towards the planet's position if the disc is marginally optical thick. The envelope emitting region can be extended and elongated, since
the Hill radius can be a moderate fraction of the disc scale height and the spirals also contribute to the radio emission.
Ultimately, to confirm the detection, we need multi-wavelength observations or/and additional signatures (e.g. spirals).

Strategically, to find these embedded planetary envelopes, we should search for disc regions that are marginally optically thin. A too small
optical depth leads to too little thermal emission, while a too large optical depth can hide the planetary envelope. Enough spatial resolution that can marginally resolve the clump and high
sensitivity are desired. Multi-wavelength observations can also help us to confirm the detection and reveal the spectral index in the envelope.

Finally, our model suggests that the clump detected by ALMA at 52 au in TW Hydrae disc
is consistent with the envelope around an embedded 10-20 $M_{\oplus}$ planet. The observed flux, the spectral index dip, and the tentative spirals are all consistent with our embedded planet model. Since the planet is embedded in a dusty disc with mm-sized pebbles, the planet may be accreting 
pebbles from the disc. The derived pebble accretion rate can double the core mass over 1 Myrs. The luminosity from the pebble accretion is comparable to the luminosity from KH contraction, and is also consistent with observations. {  With a moderate ALMA integration time, TW Hydrae is the only source that we can detect the envelopes around low-mass planets. It is the closest protoplanetary disc so that its disc scale height can be resolved by more than 1 beam with enough sensitivity. It is also a face-on system, and the disc region at $\sim$50 au is marginally optically thin. Finding a low-mass planet candidate in the only system where we can potentially detect such planets implies that these low-mass embedded planets could be common in protoplanetary discs.  }
Future ALMA
and ngVLA observations may reveal these low-mass planets ($\gtrsim$8 $M_{\oplus}$) beyond 10 au, shedding light on core accretion theory (e.g. pebble accretion) and even constraining the planet formation theory through the ``protoplanet'' population.

\section*{Acknowledgments}
The authors thank the referee Eve J. Lee for a helpful and thorough review which improves the presentation of this work. 
Z. Z. acknowledges support from the National Science Foundation under CAREER Grant Number AST-1753168.

\section*{DATA AVAILABILITY}
The data underlying this article will be shared on reasonable request to the corresponding author.




\bibliographystyle{mnras}
\bibliography{Bibliography}
\bsp
\label{lastpage}
\end{document}